\documentclass[aps,prd,reprint,superscriptaddress]{revtex4-1}
\usepackage{amsmath,amssymb}
\usepackage[colorlinks=true,allcolors=blue]{hyperref}
\usepackage{dcolumn}
\usepackage{units}
\usepackage{graphicx}
\usepackage{float}
\usepackage{enumitem}
\usepackage{diagbox}
\usepackage{overpic}
\usepackage{wasysym}
\usepackage{comment}
\usepackage{color}
\usepackage{longtable}

\allowdisplaybreaks[1]

\newcommand{\be}{\begin{equation}}
\newcommand{\ee}{\end{equation}}
\newcommand{\tref}[1] {Table~\ref{#1}}
\newcommand{\eref}[1] {Eq.~(\ref{#1})}
\newcommand{\erefs}[1] {Eqs.~(\ref{#1})}
\newcommand{\sref}[1] {Sec.~\ref{#1}}
\newcommand{\fref}[1] {Fig.~\ref{#1}}

\newcommand{\ve}[1]{\mathbf{#1}}

\begin{document}
\title{Testing General Relativity in the Solar System: \\ present and future perspectives}  
\author{Fabrizio De Marchi}
\email[Corresponding author: ]{fabrizio.demarchi@uniroma1.it}
\affiliation{Department of Mechanical and Aerospace Engineering, Sapienza University of Rome,   Via Eudossiana, 18, 00184 Rome, Italy}
\author{Gael Cascioli}
\affiliation{Department of Mechanical and Aerospace Engineering, Sapienza University of Rome,   Via Eudossiana, 18, 00184 Rome, Italy}


\begin{abstract}
The increasing precision of spacecraft radiometric tracking data experienced in the last number of years, coupled with the huge amount of data collected and the long baselines of the available datasets, has made the direct observation of Solar System dynamics possible, and in particular relativistic effects, through the measurement of some key parameters as the post-Newtonian parameters, the Nordtvedt parameter $\eta$ and the graviton mass.\\  
In this work we investigate the potentialities of the datasets provided by the most promising past, present and future interplanetary missions to draw a realistic picture of the knowledge that can be reached in the next 10-15 years. 
To this aim, we update the semi-analytical model originally developed for the BepiColombo mission, to take into account planet-planet relativistic interactions and eccentricity-induced effects and validate it against well-established numerical models to assess the precision of the retrieval of the parameters of interest.\\ Before the analysis of the results we give a review of some of the hypotheses and constrained analysis schemes that have been proposed until now to overcome geometrical weaknessess and model degeneracies, proving that these strategies introduce model inconsistencies. Finally we apply our semi-analytical model to perform a covariance analysis on three samples of interplanetary missions: 1) those  for which data are available now (e.g. Cassini,
MESSENGER, MRO, Juno), 2) in the next years  (BepiColombo) and 3)  still to be launched as JUICE and VERITAS (this latter is waiting for the approval). \\

\end{abstract}

\maketitle


\section{Introduction}\label{intro}
Precise radio tracking between spacecrafts and Earth stations 
has become nowadays a widely used method to access Solar System dynamics.

Range, range-rate and angular (e.g. $\Delta DOR$) measurements are used, in the standard least-squares orbit determination procedure, for several purposes, 
in particular to improve  the Solar System ephemerides or to perform fundamental physics experiments.\\
The scope of this work is to analyze the most promising past, present and future planetary missions to draw a realistic picture of the achievable degree of accuracy in the description of Solar System dynamics in the next 10-15 years.
Planetary missions are by themselves multidisciplinary, carrying a variety of instruments tailored for the investigation of one or more crucial features for the characterization of planets or satellites. Typically they are aimed to the determination of the 
 gravity field, surface imaging, the study of the atmosphere (or at least the exosphere), the magnetic field, etc.
 Most missions involve orbiters or landers, rather than flyby probes, which are suited for long-lasting observation campaigns.
For the purposes of this work, 
we are interested  in the orbital part of the missions only, characterized by the capacity of a high precision determination of the position and velocity (the state vector) of the probe relative to the target body center of mass, as a result of the intensity of the gravity field, and thus by the possibility of high precision determination of the state vector of the target body itself with respect to the Solar System fundamental plane.
In this work we will consider separately seven  
interplanetary space missions.  Two 
of them (MESSENGER, Cassini) have ceased their operation in 2015 and 2017 respectively, three are currently active (Mars Reconnaissance Orbit, BepiColombo, Juno), one approved but still to be launched (JUICE). Finally, we also consider the proposed, but still not approved, Venus orbiter named VERITAS as a test case for future Venus orbiters equipped with state-of-the-art tracking systems.\\
 In the limit of "weak field", which is a good approximation for the Solar System environment, the Einstein field equations can be expressed in terms of small deviations from Newton's laws \cite{will1993}.
 More generally, { these deviations are used to be parametrised by the set of coefficients called post-Newtonian (PN) parameters. In the Newtonian theory of gravity they are all zero, but  in General Relativity (GR) some of them are, by definition, unity. Therefore, their precise measurement is the key to test GR in the Solar System.  
  The PN parameters $\alpha_1,\alpha_2$ are nonzero in case of violation of the Einstein's principle of relativity (i.e. the laws of physics are independent of the 
   {reference frame}).
  The parameters $\beta$ and $\gamma$ (in GR both equal to 1) are related, respectively, to the  degree of nonlinearity of the gravitation and to the space curvature generated by the unit rest mass. 
In addition to its dynamical role (i.e. it appears into the equations of motion), the parameter $\gamma$ controls the so-called "Shapiro delay"\cite{shapiro1964} to the light-time of a radio beam. The delay being enhanced when the radio beam passes in proximity of the Sun.\\
The description of the gravity in GR as a mere geometric effect is based on the Equivalence Principle (EP), which states that the gravitational mass is identical to the inertial one. A possible violation of the EP has been tested several times in the last centuries and up to now it has never been measured \cite{adelberger2009}.\\ 
The ''weak form'' (WEP) of the EP states the universality of the free-falling of test particles in an external field. Conversely, the ''strong form'' (SEP) generalizes this effect to bodies whose gravity fields are non-negligible (e.g. planets or stars).\\
The most widely held theory that accounts for SEP violation, relates the inertial and gravitational masses through the self-gravitational potential of the involved bodies scaled by the Nordtvedt parameter $\eta$ which is 
related to PN-parameters by the Nordtvedt equation (valid for metric theories) \cite{nordtvedt1970,will1993}
 \be
\eta=4 \bar \beta - \bar \gamma -\alpha_1 -\frac{2}{3}\alpha_2; \quad \mbox{with }\bar \beta=\beta-1; \quad \bar \gamma=\gamma-1
\label{eq:nordtvedt}
\ee
 which is derived from the assumption that the metric tensor $g_{ij}$ is symmetric.\\
Other parameters we are interested in are the GM of the Sun (hereafter $\mu_0$), its rate of change in time $\zeta=\dot \mu_0/ \mu_0$ (not to be confused with PN parameter $\zeta$),
   the gravitational oblateness of the Sun $J_{2\odot}$ and the  angular momentum of the Sun $S_\odot$ which can be inferred by observing the Lense-Thirring effect \cite{lense1918}.\\
   Finally, we  also add to the list the Compton wavelength  $\lambda_g$ of the graviton which, as proposed by \cite{will2018},  can be measured by Earth-Mars ranging.\\ 
The parameters described above produce effects on the long-term trajectories of the Solar System bodies. 
It is well known that ranging data are more suited to convey informations about long-term perturbations than range-rate, which are the main source of information about "local" and fast-changing accelerations. Generally speaking, Doppler data are used to resolve the trajectory of the probe around the host planet center of mass, while the range data are used to resolve the motion of the planet center of mass with respect to Solar System barycenter.

 For this reason, the experiments devoted to fast-changing signals (e.g. planetary gravity fields), which are based mainly on range-rate measurements,
 are in practice uncorrelated to fundamental physics ones. Therefore, we can neglect the motion of the probes around the reference planet and focus on the Earth-to-planet distance perturbation.\\
 Unfortunately, our knowledge about most of Solar System asteroids masses and positions in time is not exact. 
 Thus, assuming these quantities as perfectly known in our model would introduce spurious signals resulting in biases on the estimated  parameters and in the underestimation of their formal uncertainties.\\
 When dealing with simulated data these effects can be detected by comparing the true errors (estimated minus expected values) and the formal errors,
  but, dealing with real data, the true errors are unknown.\\
 An example: the current uncertainty of the Jupiter GM is $1.5-2.0$~km$^3$/s$^2$ \cite{folkner2017}, leading to a perturbation on the Earth-Mercury range of about 20-30mm (see \cite{demarchi2016}, Tab. IV, App. E). In \fref{fig:eta_mu5} we show this effect (red line).\\
 For comparison, in the same figure (black line) we report the Earth-Mercury range perturbation due to $\eta=10^{-4}$ (i.e. the current accuracy about $\eta$).\\
 { The two signals are very similar because the perturbation due to $\eta$ depends on the position of the Sun with respect to the Solar System Barycenter that, in turn, is determined mainly by Jupiter mass and position.}
 { Anyway, the }
 signals are 
 not identical  (the Pearson correlation coefficient is 0.87),  therefore it is still possible to measure $\eta$ with an accuracy below the $10^{-4}$ level \cite{demarchi2016,genova2018}. 
However, this also means that to further reduce the uncertainty of $\eta$ it is necessary to improve the knowledge of the Jupiter's GM. 
This is true, in general for all parameters we are interested in: precise estimations are strongly conditioned to the accuracy on the
 ephemerides and masses of the Solar System bodies.\\
\begin{figure}[h!]
\includegraphics[width=.89\columnwidth]{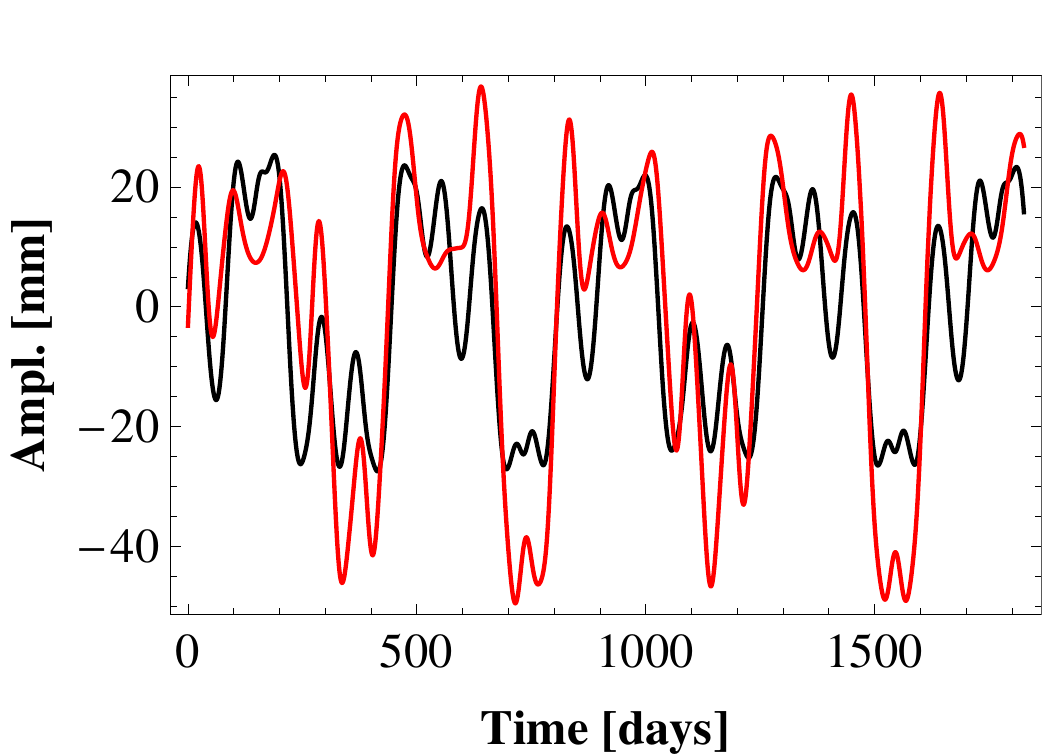}
\caption{\footnotesize Red line: Earth-Mercury spurious signal on range caused by a 1.5~km$^3$/s$^2$ biased assumption of the Jupiter's GM. Black line: Earth-Mercury range perturbation due to $\eta=10^{-4}$. 
Both signals have been calculated using the "non-homogeneous" contribution only  of the equations of motion (see \sref{app:hill} for details).}
\end{figure}
\label{fig:eta_mu5}
\\
This paper is structured as follows:  in \sref{sec:model} we describe the mathematical model we used for the covariance analysis, in \sref{sec:missions} we report details about the missions considered, in \sref{sec:discussion} we discuss the results and finally, in \sref{sec:conclusion} we draw the conclusions.\\

\section{Mathematical model}\label{sec:model}
The model was originally developed to be applied to MORE (Mercury Orbiter Radioscience Experiment) onboard BepiColombo (BC).  A detailed description of it can be found in \cite{demarchi2016}, 
 here we report on some important updating and improvements. 
 First of all the code has been generalised to model the range measurements, as a function of the set of parameters we are interested in, between two arbitrary planets of the Solar System.\\
We adopt the notation of \cite{moyer2000},  we define $\ve r_i$ as the coordinates the planet $i$ in an inertial reference frame,  
 $\ve r_{ij}=\ve r_j-\ve r_i$  as the vector from $i$ to $j$ and $r_{ij}=\vert \vert \ve r_{ij}\vert \vert$  its modulus.\\ 
 We number the planets 
  from 1 (Mercury) to 8 (Neptune), while 0 is referred to the Sun. In the following we will indicate with $i$ and $k$ the two planets, where $i$ is the nearest to the Sun.\\
 Finally, we call  
 $\ve q$ the list of  $N$ parameters involved for the covariance analysis. All $q_j$ are very small, consequently the perturbation on the $i$-to-$k$ distance can be expressed 
 as a first order Taylor series expansion around the nominal solution. To get the perturbations on the range measurements due to $q_j$ we follow this procedure
\begin{enumerate}
\item we write the forces per unit mass $\delta \ve a_i^j$ on body $i$ and $\delta \ve a_k^j$ on body $k$;
\item by solving the  Hill's equations of motion \cite{cw1960}  assuming 
 null initial conditions we obtain the secular { displacements} $\delta \ve r_i^j$ and $\delta \ve r_k^j$ and the range perturbed vector $\delta \ve r_{ik}^j=\delta \ve r_k^j-\delta \ve r_i^j$;
\item  the perturbed vector is finally projected along the  the $i$-to-$k$ direction to obtain the perturbation $\delta \rho_{ik}^j (\ve q)$ on the range. 
\end{enumerate}
Initial conditions of $i$ and $k$ belong to $\ve q$. Their signatures are computed { using} the homogeneous solution  \eref{eq:hom2} of Hill's equations.\\
All perturbations that will be treated, in the Hill's frame, are expressed as the sum of polynomial and sinusoidal (resonant and non-resonant) functions of time.  The corresponding solutions of the Hill's equations are reported in \sref{app:hill}.\\
Finally, after imposing a-priori for all parameters  based on their current or expected (in the case of future missions) accuracies, we calculate the covariance matrix (see \sref{sect:apriori}).

\subsection{Improving the model}\label{sec:model1}
First of all, we included the effect of the orbital eccentricity of Mercury since it is considerably larger ($e_1=0.205$) than those of the other planets. Moreover, we 
add to the model described in \cite{demarchi2016} the following parameters: the Eddington parameter $\gamma$, the preferred frames parameters ($\alpha_1,\alpha_2$), the angular momentum  of the Sun ($S_\odot$) 
 and the Compton wavelength of the graviton ($\lambda_g$). Finally, we also consider the effects of the uncertainties on the positions of planets and asteroids.\\
The model does not include the $z$-coordinates since orbital inclinations of the most massive bodies are in general very small.\\
For validation purposes, the ranging signatures obtained with the model presented in this work have been compared to numerically integrated ones. 
The results in the case of a simulated Earth-Mercury ranging experiment are reported in \fref{fig:partials}.\\ The analytical model shows to be in good agreement with the numerical solutions presenting small deviations, due mainly to our co-planar approximation. These small deviations (that are in the order of 15-20 \% in terms of relative error) map into negligible differences in terms of variance and covariance on the solve-for parameters. More specificly, it can be proven that a relative error $e_r$ on the range signatures maps into an error of the order of $e_r$ (in a worst-case assumption corresponding to an additive and constant error) on the RMS, meaning that, for the purposes of this work, the approximation introduced by the co-planar assumption is justified.\\
 The signatures of some parameters in the ranging data, for the 5 different ranging options here considered (Earth-Mercury, Earth-Venus, Earth-Mars, Earth-Jupiter, Earth-Saturn), are displayed in \fref{fig:signatures}.

\subsubsection{Mercury's eccentricity}\label{sub:mercuryecc}
 In our co-planar approximation we will use the longitude of the perihelion $\varpi=\Omega+\omega$ where $\Omega$ is the  longitude of the ascending node and $\omega$ is the argument of perihelion.\\
The heliocentric cartesian position of body $i$ is 
 $\ve r_{0i}=\{x_{0i},y_{0i}\}$, where 
\be
\begin{split}
x_{0i}=&  r(t) \left[ \cos \nu(t) \cos \varpi- \sin \nu(t) \sin \varpi \right];\\
y_{0i}=&  r(t) \left[ \sin \nu(t) \cos \varpi+ \cos \nu(t) \sin \varpi \right].
\label{eq:r01}
\end{split}
\ee
 where $\nu$ is the true anomaly. We define $R_{0i}$ the semimajor axis of body $i$,  $t^p$ as the epoch of the passage at the perihelion and $t'=t-t^p$.\\
The terms $(r \cos \nu, r \sin \nu)$ in \eref{eq:r01} at the first order of the elliptical expansion are  \cite{brouwer1961} 
\be
\begin{split}
r \cos \nu &= R_{0i} \cos (n_i t') + e_i \frac{R_{0i} }{2} \left[\cos (2 n_i t')-3 \right] + O(e_i^2);\\
r \sin \nu &= R_{0i} \sin (n_i t') + e_i \frac{R_{0i} }{2} \sin (2 n_i t')  + O(e_i^2);
 \end{split}
\label{eq:ecc1}
\ee
where $ n_i=\sqrt {(\mu_0+\mu_i)/R_{0i}^3}$ is the mean motion.\\
The accuracy of \erefs{eq:ecc1}, and consequently of the model in general, is $O(e_i^2)\approx 4\%$ (for Mercury), which is sufficient for our purposes.\\
Referring to the circular orbit approximation, we define the radial/ transverse unit vector
\be
\begin{split}
\ve u_r^i &= \{ \cos (\Phi_i+\varpi_i) , \sin (\Phi_i+\varpi_i), 0 \}; \\
\ve u_t^i &= \{-\sin (\Phi_i+\varpi_i) , \cos (\Phi_i+\varpi_i), 0 \}; 
\end{split}
\ee
where
\be
\Phi_i= n_i (t-t_i^p).
\label{eq:phi}
\ee
Finally, from \eref{eq:r01}, the vector $\ve r_{0i}$, projected along these directions, is  \cite{demarchi2012}
\be
\ve r_{0i} \approx R_{0i} \ve u_r^i + e_i \ve R_i^e
\label{eq:r0j}
\ee
where 
\be
\ve R_i^e = R_{0i} (-\cos \Phi_i \ve u_r^i + 2 \sin \Phi_i \ve u_t^i)
\ee
is the first-order correction for eccentricity.\\
The formulas necessary to express the forces described in this Section in terms of the Mercury's orbital eccentricity are reported in  \sref{app:ecc}.

\subsubsection{Planetary perturbations on range}\label{sub:plpertrange}
In \cite{demarchi2016} only the perturbations due to uncertainties of GMs were considered, for completeness we included also the uncertainties on the positions of the bodies.\\
Define $\ve p$ a vector containing the orbital parameters and the GM of a planet/asteroid. It can be expressed as $\ve p= \bar {\ve p} + \delta \ve p$,  
   where $\bar {\ve  p}$ are the adopted values (e.g. from an arbitrary set of ephemerides) and $\delta \ve p$ are  small deviations between true and the adopted values.\\ 
Orbits are assumed co-planar and the eccentricity of Mercury is assumed to be known, therefore the elements of $\ve p$ are: mean motion, 
 radius ($R_{0j}$) of the orbit, the GM ($\mu_j$) and the longitude of the perihelion ($\varpi_j)$. 
 We considered $N=353$ perturbers among planets and asteroids, so the parameters to be included into the original sample are $3\times N$.\\
The  trajectory of body $i$, in heliocentric coordinates, can be obtained by solving
\be
\ddot { \ve r}_{0i} = -\dfrac{\mu_0+\mu_i}{r_{0i}^3}\ve r_{0i} + \displaystyle \sum_{j \neq i \neq 0} \mu_j \left(\dfrac{\ve r_{ij}}{ r_{ij}^3} -  \dfrac{\ve r_{0j}}{ r_{0j}^3}\right).
\label{eq:r0i}
\ee
Since $\mu_j \ll \mu_0$ the deviation from the Keplerian orbit is small (in the interval of time of our interest) and the  term into summation can be calculated 
  by assuming unperturbed orbits for each $j$.\\
As said above, we assume $e_j=0 \,\forall\, j \neq 1$.\\ 
\eref{eq:r0i}  can be re-written in the Hill's frame  \cite{cw1960}  by using \eref{appb} and \eref{eq:pl} (with $n=3$) and solved for 
planets $i$ and $k$ (see \sref{app:hill} and \cite{demarchi2016} for details) finding the analytical expression of $\ve r_{ik} (\delta \ve p)$.\\
Expanding at the first order this quantity and summating over all bodies, the perturbations are
\be
\delta \ve r_{ik}= \sum_{j\neq i\neq k} \left[ \delta \mu_j \dfrac{\partial \ve r_{ik}}{\partial \mu_j}\bigg\vert_{\bar \mu_j} + \delta R_{0j} \dfrac{\partial \ve r_{ik}}{\partial R_{0j}}\bigg\vert_{\bar R_{0j}}+\delta \varpi_j \dfrac{\partial \ve r_{ik}}{\partial \varpi_j}\bigg\vert_{\bar \varpi_j}\right].
\ee
Finally, the range perturbation is
\be
\delta \rho_{ik} \approx \frac{\delta \ve r_{ik} \cdot \ve R_{ik}}{R_{ik}}
\label{eq:drho13}
\ee
and the factor $1/R_{ik}$ can be obtained using \eref{eq:leg}.

\subsubsection{Dynamical effects of $\gamma$  and other parameters}\label{sec:gamma}
Defining $ \mu_0$  as the current ($t=t_0$) estimated value of the GM of the Sun, we can write the true GM as
\be
\mu_{0, true}(t)=  \mu_0 \left[1+ \zeta (t-t_0) \right]+ \delta_{\mu_0}
\ee
where $\mu_0$ is the current best estimation and $\delta_{\mu_0}$ is the offset.\\
We also re-consider the perturbation due by the gravitational flattening of  the Sun. In \cite{demarchi2016} the inclination of the Sun's equator was included, 
 however its effect is smaller than the one triggered by the eccentricity of the Mercury's orbit. Therefore we decide to neglect the first effect and include the second one. \\
Neglecting planet-planet interactions, the force per unit mass on body $i$ due to $\bar \gamma$, $\bar \beta$, $\delta_{\mu_0}$, $\zeta$ and  $J_{2 \odot}$ is  \cite{will1993}\\
\be
\begin{split}
\delta \ve a_i^{\bar \gamma, \bar \beta, \delta_{\mu_0}, \zeta, J_{2 \odot}} = \frac{\mu_0}{r_{0i}^3} \left[ -\left(\zeta t +\frac{\delta_{\mu_0}}{\mu_0} \right) {\ve r_{0i}}  +2 (\bar \beta+\bar \gamma) \frac{\mu_0}{c^2} \frac{\ve r_{0i}}{r_{0i}}+\right.\\
\left. - J_{2\odot}\frac{3}{2}   \frac{R_\odot^2 }{r_{0i}^2} \ve r_{0i}- \bar \gamma \frac{\dot r_{0i}^2}{c^2} {\ve r_{0i}}+2 \bar \gamma \frac{\ve r_{0i} \cdot \dot {\ve r}_{0i}}{c^2} {\dot {\ve r}_{0i}}   \right] .
\end{split}
\label{eq:da}
\ee
 where  $R_\odot$ is the Sun's radius.  
The terms into \eref{eq:da} can be written in the Hill's frame  by using  \eref{eq:appecc}. \\
Again, neglecting planet-planet interactions, for preferred frames parameters the perturbation on the orbit of body $i$ is  \cite{will1993}
\be
\begin{split}
\delta \ve a_i^{\alpha_1,\alpha_2} = \frac{\mu_0}{2 c^2 r_{0i}^3} \left[ \alpha_1 (\ve w \cdot \dot {\ve r}_{0i}) \ve r_{0i}+3 \alpha_2 (\ve w \cdot \ve r_{0i})^2 \frac{\ve r_{0i}}{r_{0i}^2} +\right.\\
\left. -\alpha_1 (\ve r_{0i} \cdot \dot {\ve r}_{0i}) \ve w- 2 \alpha_2 (\ve w \cdot \ve r_{0i}) \ve w + w^2 (\alpha_1-\alpha_2) \ve r_{0i} \right]
\end{split}
\label{eq:da2}
\ee
where $w^2=\ve w \cdot \ve w$ and
\be
\ve w = 3.69 \times 10^5 \{-0.970  , 0.139 , -0.197 \}   \quad \mbox{[m/s]}
\ee
 is the velocity of the Solar System Barycenter (SSB) relative to the thermal background radiation \cite{durrer2015} that represents the ''preferred frame''.\\
 It can be seen from \eref{eq:da} and \eref{eq:da2} as the eccentricity of the orbit of body $i$ reduces the degeneracy among parameters.

  \subsubsection{Lense-Thirring effect}
 The nonzero angular momentum of the Sun produces a secular precession of  both the longitude of the ascending node and the argument of perihelion  of a planet. It is a relativistic correction called "Lense-Thirring effect" \cite{lense1918}. 
 The perturbation on body $i$  is \cite{moyer2000}
  \be
  \delta \ve a_{LT}^i = G S_\odot  \frac{\gamma+1}{c^2 r_{0i}^3} \left[-\ve s \times \dot {\ve r}_{0i}+3 \frac{(\ve s \cdot \ve r_{0i}) (\ve r_{0i} \times \dot {\ve r}_{0i})}{r_{0i}^2} \right],
  \label{eq:LT}
  \ee
where $S_\odot=1.92 \times 10^{41}$[kg m$^2$s$^{-1}$] \cite{iorio2012} is the modulus of the angular momentum of the Sun and $\ve s$ is the unit vector which indicates its direction.\\
We  neglect the second term into square brackets because the cross product is a vector along the $z$-axis and in our co-planar approximation it cannot be detected. The term $(\ve s \cdot \ve r_{0i})$ is also very small because $\ve s$ is near aligned along $z$-axis.\\
We express the angular momentum of the Sun as $S_\odot=k_{LT} M_\odot R_\odot^2 \Omega_\odot$ where $\Omega_\odot=2.864 \times 10^{-6}$ [rad/s] is the angular velocity  at the equator \cite{bertotti1993} and $k_{LT}$ the normalized moment of inertia (i.e. the parameter to be estimated).
Assuming $\gamma=1$, after some algebra we get 
 \be
  \delta \ve a_{LT}^i \approx  k_{LT}   \frac{2 \mu_0  R_\odot^2 \Omega_\odot}{c^2}  \left[\frac{\dot {\ve r}_{0i}}{r_{0i}^3}\times \{0,0,1\} \right].
  \label{eq:LT2}
  \ee
 The expression in the Hill's frame of the term into square brackets of \eref{eq:LT2}  is reported in \eref{appf}.
\subsubsection{Shapiro delay}
The Shapiro delay \cite{shapiro1964} is one of the consequences of  the space-time curvature in the proximity of the Sun. 
It affects both range and Doppler data and it can be used to  measure  the parameter $\gamma$ in an independent way from the dynamical effect described in \sref{sec:gamma}.\\
Regarding the signature on the range, the following  term \cite{moyer2000}
\be
\delta r_{ik}^{\bar \gamma}=\ \bar \gamma  \frac{\mu_0}{c^2} \log \frac{r_{0i} + r_{0k}  + r_{ik}} {r_{0i}+ r_{0k} - r_{ik}}
\label{eqshapiro}
\ee
 must be added to the dynamical effect of $\bar \gamma$ ($i$ and $k$ are the inner and the outer planets, respectively).\\
 From \eref{eqshapiro} one can notice as the signal is maximized during the $i$-Sun-$k$ alignments. They   can occur during the cruise phases as well during the orbital  phases. For probes orbiting around inner (outer) planets 
 the alignments correspond to superior conjunctions (oppositions).\\
 The perturbation in \eref{eqshapiro} can be evaluated with good approximation (for our purposes) by using the formulas \sref{app:ecc}.

\subsubsection{Compton wavelength of the graviton}
The gravitational potential for a system of bodies in the case of a massive graviton is  \cite{will2018}
\be
U= - G \sum_{i\neq j} \frac{m_i m_j}{r_{ij}}\, e^{-r_{ij}/\lambda_g}
\ee
where $\lambda_g$ is the {\em Compton wavelength} of the graviton which, in the classic limit, is infinity.\\ 
The previous lower limit  $\lambda_g > 2.8 \times  10^{12}$~km (obtained by measurements based on solar-system dynamics  \cite{talmadge1988}), has been  increased to   $\lambda_g > 1.6 \times 10^{13}$~km thanks to direct detection of gravitational waves (three events GW150914, GW151226 and GW170104) \cite{bernus2019}.
 Recently, \cite{will2018} forecasted that the lower limit of $\lambda_g$ could be increased to  1.2--2.2 $\times$ 10$^{14}$~km thanks to precise measurements of the  perihelion advance of Mars obtained from MRO data.\\
   In this work we will check this prediction taking into account also the correlations of $\lambda_g$ with  all other parameters.
Neglecting the planet-planet interaction, the force per unit mass  on planet-$i$ is
\be
\ve a_i =  -\frac{\mu_0}{r_{0i}^3} \left[1- \frac{r_{0i}^2}{2 \lambda_g^2} \right] \ve r_{0i}+ O(1/\lambda_g^3).
\label{eq:gr}
\ee
Therefore, using \eref{appb}  with $n=1$, the perturbation due to the massive graviton is
\be
\delta \ve a_i ^{\lambda_g}=   \frac{\mu_0}{2 \lambda_g^2} (\ve u_r^i+2 e_i  \sin \Phi_i \ve u_t^i)
\label{eq:gr2}
\ee
The (small) parameter belonging to $\ve q$ is  $1/\lambda_g^2$ and its formal error, say $\sigma_g$, extracted  from the covariance matrix, gives the lower limit to the
 Compton wavelength of the graviton
 \be
 \lambda_g > 1/\sqrt{\sigma_g}.
 \ee

As an example, we plot in \fref{fig:comparison} the signatures on the Earth-Mercury range due to $\bar \gamma=10^{-5}$ and $\lambda_g=10^{14}$~km. We compare the ones obtained by solving the Hill's equations (with and without eccentricity of Mercury) with those obtained by numerical integration. The contribution of the eccentricity of Mercury's orbit is evident.
\begin{center}
\begin{figure}[h!]
\includegraphics[width=.9\columnwidth]{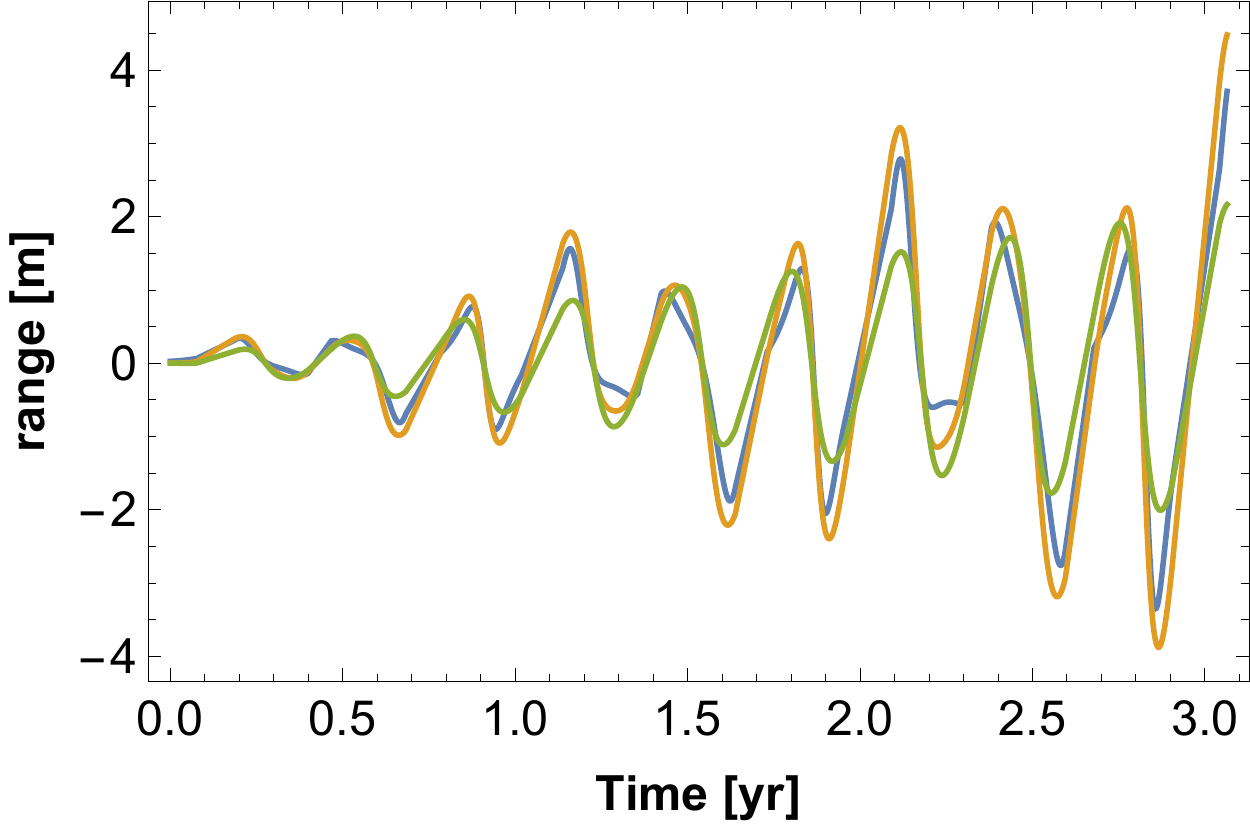}
\includegraphics[width=.9\columnwidth]{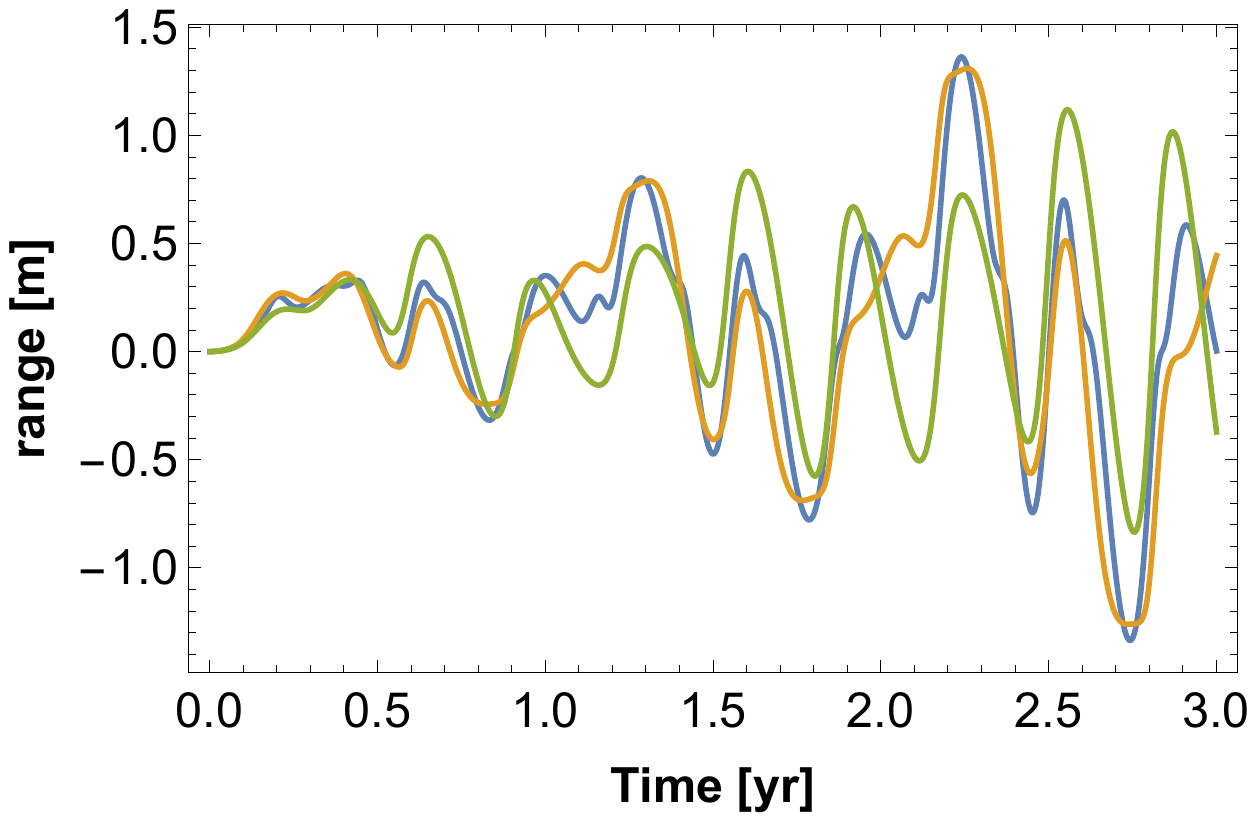}
\caption{\footnotesize Signature on the Earth-Mercury range due to $\bar \gamma=10^{-5}$ (top) and $\lambda_g=10^{14}$~km (bottom). Blue: numerical integration, green: analytical model (assuming $e=0$), orange: analytical model with first order eccentricity of Mercury's orbit included.}
\end{figure}
\label{fig:comparison}
\end{center}

\subsubsection{Non-gravitational forces}\label{sect:nong}
 Non-gravitational forces, for our purposes, must be treated as a  source of systematic noise.  
They are in general non-deterministic (i.e.  solar pressure or thermal effects) and must be calibrated or,  
 thanks to the EP,  directly measured by an onboard accelerometer (it is the case of BC) and subtracted to the dynamical model.
For the mission JUICE, the accelerometer will be necessary to measure the displacement of the onboard antenna with respect to the center of mass due to the sloshing of the propellant \cite{cappuccio2018a}.\\
The calibration of the non-gravitational forces, usually composed by both stochastic and deterministic signals, is commonly tackled with a multi-arc strategy (see \cite{MG} for a detailed description).\\
In this work, for simplicity, we will take into account these effects in an indirect way, by assuming conservative values for the RMS of the single measurements. In particular we considered a value of RMS associated with the measurements that is always a factor 2 or 3 larger than the observed (or predicted, in the case of future missions) value.

\subsubsection{Aging of the transponder}\label{sect:aging}
The on-board transponder develops a cumulative offset that corresponds to a drift in the range signal. This phenomenon is called "aging" of the transponder. 
 In the case of MESSENGER (hereafter MSG), a linear trend leading to an offset of $<1$m over 7 years, attributed to the aging of the transponder, 
 has been observed and removed from the radiometric data \cite{genova2018}.
For BC,  the onboard  self-calibrating system is expected to reduce the aging to less than few cm after 1 year,  this effect can be absorbed by a monotonic function of the time \cite{schettino2016}. 
Therefore, we added  the following effect to the range signals
\be
\delta \rho_{ik}  = \rho_{start}+\Delta \rho \frac{t- t_{start}}{t_{end}-t_{start}}
\label{eq:aging}
\ee
 and ($\rho_{start},\Delta \rho$) are  parameters to be estimated.

\subsection{Covariance analysis and a-priori}\label{sect:apriori}
Given the set of $N$ parameters $\ve q$, our purpose is the calculus of their RMS, which are the  square roots of the diagonal elements of the  covariance matrix.
We will take into account also the current knowledge about some (or all) parameters, so we will use a-priori, that can be seen as additional "observations" \cite{MG}.\\
 The simplest way is to express them as a set of $M$ 
 linear combinations of elements of $\ve q$ 
 \be
 f_n^P = \sum_{j=1}^N a_{nm} (q_m- q_{m,true}) ;\quad n=1,...,M
\ee
where $a_{nm}$ are constant coefficients and $f_n^P$ are normally distributed with zero mean and variance $\sigma^P_n$  which is the accuracy of the a-priori information.\\
The normal and covariance matrices are
\be
\ve C = \ve H^T \ve W \ve H + (\ve H^P)^T \ve W^P \ve H^P; \qquad \ve P = \ve C^{-1}
\ee
where $\ve H$ and $\ve H^P$ are the design matrices

\be
\begin{split}
H_{lm} &= \frac{\partial \rho_{ik}(t_l) }{\partial q_m}; \quad l=1,...,N_d; \quad  m=1,...,N;\\
H^P_{nm} &= \frac{\partial f_n }{\partial q_m}; \quad m=1,...,N; \quad n=1,...,M\\
\end{split}
\ee
where $N_d$ is the number of  observations (i.e the $i$-to-$k$ range measurements and  $\ve W,\ve W^P$ are weight matrices, here assumed to be diagonal

\be
\begin{split}
W_{lm}&=\frac{\delta_{lm}}{\sigma_l^2}; \quad l,m=1,...,N_d; \\
W^P_{mn} &=\frac{\delta _{mn}}{(\sigma_n^P)^2}; \quad m,n=1,...,M.
\end{split}
\ee
with $\delta_{ij}$ being the Kroneker delta.\\ A-priori have been assumed according to the current knowledge of the parameters involved.\\
For planets, the current uncertainties about GMs and ephemerides  are reported in \tref{tab:status2}, the last ones being based on the comparison among recent ephemerides DE430, INPOP15a and EPM2014 \cite{IAU2016}. We used these values for planetary $\sigma(\mu_j),\sigma(R_{0j}),\sigma(\varpi_j)$.  Thanks to the data carried by Juno mission, in the next future the GM of Jupiter will be likely reduced to $\sigma(\mu_5)=1.0$~km$^3/s^2$ \cite{notaro} therefore we adopt this value.\\
Among the 345 minor bodies we considered, two are  trans-neptunians (Pluto, Eris) for which we used (500km, 20mas) for radial/along-track uncertainties. All other bodies are asteroids belonging to the main belt and we assumed a  positioning error of (100km, 10mas).

\begin{table}[!h]
\begin{ruledtabular}
\caption{Current knowledge about GMs and positions of the Solar System planets.}
\begin{tabular}{llll}
                                                  &   Radial                                   &   along track              & GM   \\ 
body                                           &    position                                  &   position             & unc.   \\   
                                                   &  unc. [km]                               &  unc.    [mas]                &    [km$^3$/s$^2$]     \\   
 \hline                  
Mercury               &  0.020             &          0.2              &  0.9     \cite{mazarico2014}       \\ 
Venus                  &  0.004             &          0.2              &   0.006 \cite{luzum2011}      \\  
Earth                  &   0.002              &          0.2             &    0.0043  \cite{luzum2011}   \\  
Mars                    &   0.002             &          0.2             &    0.00028  \cite{luzum2011} \\  
Jupiter                  &   1.5              &           4.0             &     2.0 \cite{folkner2017}--1.0 \cite{notaro}            \\  
Saturn                 &    0.2               &           0.2             &     1.1        \cite{luzum2011}   \\  
Uranus                &    50.               &          5.0              &     7.0         \cite{luzum2011}  \\  
Neptune               &   200.             &           5.0             &     10.0        \cite{luzum2011} \\  
 \end{tabular}
 \label{tab:status2}
\end{ruledtabular}
\end{table}
To summarize, a-priori have been used for
\begin{enumerate}
\item  Parameters related to planets and asteroids: 
 $\mu_j$, $R_{0j}$ and $\varpi_j$;
\item  Nordtvedt equation (\ref{eq:nordtvedt}).  Being an exact relation, in practice it corresponds to express one of the parameters as a function of the others. This can be done by  adopting a small, but nonzero, value for $\sigma_n$. Following   
\cite{schettino2018}, we put $\sigma_n^P=10^{-8}$;
\item initial state vectors for planets $i$ and $k$  based on the values into \tref{tab:status2}  (see \sref{sec:ciniz} for a detailed description);
\item parameters   $\beta$, $\gamma$, $\eta$, $\alpha_1$, $\alpha_2$, $\mu_0$, $J_{2\odot}$,  $\zeta$, $k_{LT}$ and $\lambda_g$ using as a-priori reported in \tref{tabapriori}.
\end{enumerate}

\subsubsection{A-priori on initial conditions}\label{sec:ciniz}
 Initial conditions can be expressed in barycentric ($\ve r_i, \dot {\ve r}_i,\ve r_k, \dot {\ve r}_k $) as well as heliocentric ($\ve r_{0i},\dot {\ve r}_{0i},\ve r_{0k},\dot {\ve r}_{0k}$) coordinates.\\
  In our model, the natural setup is  the heliocentric one. The evolution of the state vector in terms of the initial conditions is given by \eref{eq:hom2} where initial conditions  $\{r_0, t_0,\dot r_0,\dot t_0\}$ are relative to the Hill's frame (they are related to heliocentric coordinates by \eref{eq:app1} evaluated at $t=0$).\\
  We can pass to barycentric coordinates by using
  \be
  \ve r_i = \ve r_{0i}- (1 + \eta \Omega_0 ) \frac{\sum_{j \neq 0} \mu_j \ve r_{0j}}{\sum_j \mu_j} + O(\eta^2)
  \label{eq:ciniz}
  \ee
{where $\Omega_i$ is the ratio between self-gravitational and rest energy of body $i$ (for the Sun $\Omega_0=-3.52 \times 10^{-6}$) \cite{demarchi2016}.\\
From \eref{eq:ciniz}, the passage to a barycentric configuration implies the introduction of a signal proportional to $\eta$, to be added to the "heliocentric" one which is obtained by solving the equation of motion
\be
\ddot {\ve r}_{0i} = -\frac{\mu^\star}{r_{0i}^3}\ve r_{0i}+\eta \sum_{j\neq i \neq 0} \mu_j \left( \Omega_i \frac{\ve r_{ij}}{r_{ij}^3}- \Omega_0  \frac{\ve r_{0j}}{r_{0j}^3}\right)
\label{eq:eta}
\ee
where
\be
\mu^\star=\mu_0+\mu_i+\eta (\Omega_0 \mu_i+\Omega_i \mu_0),
\ee
 see \cite{demarchi2016} for details.\\
It results that the "barycentric" signal largely dominates the heliocentric one, of about a factor 10 (for a comparison see top panels of \fref{fig:signatures}).\\
Unfortunately, this advantage cannot be used: an hypothetical exact knowledge about one or more barycentric initial conditions is equivalent to an information about the real position of the Sun. However,  this latter depends on  the inertial masses of the Solar System bodies, therefore on $\eta$, so we created a self-referential loop.\\
The Sun-SSB distance is  $\vert \vert \ve r_0 \vert \vert \approx 7-8 \times 10^{5}$~km and the current uncertainty on the Nordtvedt parameter is $\Delta \eta \approx 10^{-4}$. The uncertainty on the {\em real } position of the Sun due to a possible SEP violation is
\be
\Delta \eta\,\Omega_0  \vert \vert \ve r_0 \vert \vert \approx 1\, \mbox{m} .
\ee
We verified that this is a qualitative "dividing line" for the a-priori on initial positions of bodies $i$ and $k$. If they are larger, the resulting formal error of $\eta$ is near the same in both barycentric or heliocentric configurations. In the other case, while in the heliocentric setup the RMS of $\eta$ remains stable, in the other configuration it becomes about 1 order of magnitude smaller.\\
Since the output must be independent of the configuration adopted, we conclude that the descoping strategy (the assumption of exact knowledge of some elements of the initial state vector  \cite{milani2002}, \cite{demarchi2016}, \cite{schettino2018}), when applied to a barycentric setup, leads to an unphysical RMS of $\eta$.\\
For the same reason, in this work we do not adopt constraints on rotations and rescaling \cite{milani2002}, \cite{schettino2018}, \cite{demarchi2016}  and \cite{demarchi2018}.\\
A-priori on initial conditions are sufficient to lock the rotation around $z$-axis and 
the constraint on the rescaling, as was for the descoping strategy, implicitly adds information about the real position of the Sun with respect to the SSB.} 
\\
Moreover, we retain that the rescaling is in fact not necessary since 
the astronomical unit has been redefined as a conventional unit of length (resolution B2, IAU XXVIII General Assembly 2012) and the GM of the Sun is now among the parameters to be estimated.\\
In \cite{schettino2018} it has been argued that the formal accuracies of the parameters depend on the reference epoch at which the initial state vector is estimated.\\
We believe that this effect is a consequence of erroneously maintaining the same set of a-priori on constraints (such as descoping or rescaling) when changing the estimation epoch. In  \sref{app:constraint} we prove that, if the a-priori matrix is properly propagated by using the state transition matrix $\Phi(t_1,t_0)$ the formal uncertainties remain unchanged.

  \section{Missions and datasets}\label{sec:missions}
  
Here we report an overview of the main characteristics of the interplanetary missions we considered. We use for our analysis a set of simulated "normal points" with associated errors.\\
The points with an impact parameter $b<b_{min}$ must be discarded
 because of the effect of the solar plasma noise. In the case of missions able to establish a coherent multifrequency radio link a nearly complete suppression of the plasma noise can be achieved \cite{bertotti1993}. Notably BC
 and JUICE are (or will be) equipped with a Ka-band Transponder that enables a X/X (7.2 GHz uplink / 8.4 GHz downlink), X/Ka (7.2 GHz / 32.5 GHz) and Ka/Ka (34 GHz / 32.5 GHz) band multifrequency link. 
For these missions 
we will adopt $b_{min}=7 R_\odot$, in agreement with the latest numerical simulations performed for BC \cite{imperi2018}. For the missions employing a single frequency X-band two-way link (all the others except VERITAS) we discard all the data collected for $b<b_{min}=73.5 R_\odot$ corresponding to a Sun-Earth-Probe ($SEP$) angle of $20^{\circ}$.\\
Finally, the Ka/Ka link is not expected for VERITAS, so only a partial calibration of the plasma noise will be possible. Therefore, we will adopt an intermediate value of $b_{min}=40 R_\odot$ that can be translated in a $SEP$ of about $10.5^{\circ}$ \\
The precision of each normal point depends on: 
\begin{itemize}

\item The real data sampling: $n$ points can be equivalent to one point  with RMS rescaled by a factor  $1/\sqrt{n}$ or equivalently, the measurement RMS can be rescaled to different integration times as:
\be
\sigma_\rho' = \sigma_\rho \sqrt{\frac{T_C'}{T_C}} 
\ee 
where $\sigma_\rho$ is the range RMS, $T_C$ is the integration time and $(\cdot)'$ denotes rescaled quantities.
\item The radio-link technology: a standard X-band ranging system can provide ranging measurements with an accuracy limited to a few meters, a regenerative pseudo-noise ranging system, employed on BC, VERITAS and JUICE, can reach an absolute precision of some tenths of centimeters \cite{Simone2008},\cite{imperi2018}. \\
 
 \end{itemize}
  The code we developed can deal only with Earth-to-planet range data, so particular attention must be applied when assigning the RMS of the Cassini and JUICE simulated normal points since neither Cassini nor JUICE are placed on circular orbits around Saturn or Jupiter, respectively, for the entire duration of the missions. Orbits are in general highly elliptic with several flybys of the satellites. This implies that the uncertainty in the positioning of the central body may, and in general does, vary along the duration of the mission. Specific assumptions have been made for these two missions, in order to deal correctly with this aspect (see below for details). 
 \\ 
 
 In the following we will describe case-by-case the generation of the simulated normal points.
\begin{enumerate}
\item  The scientific objectives of MSG were the study of the geology, geophysics, exosphere and magnetosphere of Mercury. It 
  has been launched in 2004. After a flyby encounter of Earth (2005), two of Venus (2006-2007) and three of Mercury (2008-2009) it began an
 in a highly elliptical orbital phase (200km x 15000km altitude, period: 12 hrs) around Mercury on March, 2011 until crashing on the surface 4 years later, on April 30, 2015.\\
The spacecraft tracking is based on a two-way X-band link as described in \cite{Srinivasan2007}. We simulated a 4.1 yrs mission, a data spacing of 10 h and an associated RMS of 1m, that is conservatively compliant with the measured performances of MSG \cite{genova2018}.

 \item Cassini was a NASA mission devoted to the study of Saturn and its environment. Launched on Oct. 15, 1997, it started the orbital phase around Saturn on July 1$^{st}$, 2004. After several flybys around the Saturn's satellites (and a 4-years orbital phase around Titan with the deployment of the lander Huygens) it  performed close passes by the rings of Saturn until the final plunge into Saturn's atmosphere on Sept. 15, 2017. Although equipped with a Ka-band transponder, during the Saturnian system tour Cassini was tracked only through a dual-frequency link (X/X, X/Ka) that does not allow a complete plasma noise cancellation scheme. For this reason we considered a $b_{min}$ corresponding to a $SEP=20^{\circ}$. We simulated a 13.2 yrs mission and,  for the accuracy of its reconstruction of the Earth-Saturn distance, we consider two cases: 100~m (optimistic) and 1~km (conservative) with a 24h sampling time.

 \item MRO (Mars Reconnaissance Orbiter)  is a still operating NASA mission devoted to the study of Mars. It has been launched on 2005 
 and the orbital insertion around Mars was accomplished on March, 2006. The orbit is near circular (250km x 316km of altitude).The tracking system operates with a single-frequency two-way X-band link and employs a DSN standard sequential tone ranging scheme \cite{Zuber2007}, that allows a ranging accuracy of the order of few meters over typical integration times (5 mins). We simulated a 12 yrs mission with a data spacing of 10 h and an RMS of 1m.
 \item Juno is a currently operating NASA mission devoted to the study of Jupiter. It entered in a polar, highly elliptical (75600km x 8.1 million km), orbit around Jupiter on July, 2016. 
 As for Cassini, 
 the
  radio-link is established through X/X, X/Ka-band links. 
 We simulated a 4.9 yrs mission with a data spacing of 53 d and an RMS of 50m. These values are motivated by the highly elliptical orbits of Juno, that allow the precise determination of Jupiter position only at perijoves, that occur every 53d. The value of the RMS reflects the current uncertainties in the reconstruction of Juno position \cite{durante}.
 
\item  BepiColombo (hereafter BC), an ESA-JAXA mission, has been  launched Oct. 19, 2018 and it will enter into Mercury orbit on Dec., 2025. After the orbit insertion there will be the release of two s/c. 
The first one, the Mercury Magnetospheric Orbiter (MMO) 
 will study of the exosphere and the magnetosphere. The other one, the Mercury Planetary Orbiter (MPO),  will be placed  on a low altitude polar orbit (480km x 1500km altitude, 2.3h period). Among the purposes of MPO there are the improvement of the gravity field and rotational state of Mercury (from MSG) and an experiment (MORE) devoted to testing GR theory.
The tracking of the spacecraft is ensured by high accuracy X/X, X/Ka and Ka/Ka band range and Doppler links between the MPO spacecraft and DSN/ESA stations. 
We generated synthetic normal points for a nominal mission of 2 yrs with a data spacing of 10 h and an associated RMS of 4 cm. This assumption is conservatively compliant with the expected performances of the BC ranging for which performances of about $1.5 \times 10^{-4}$ cm/s (at 1000s integration time) and 15cm in range (at 300s integration time) are expected \cite{iess2001}.

\item JUICE (JUpiter ICy moons Explorer) is an ESA space mission focused on the study of the icy moons of Jupiter \cite{cappuccio2018a}. The launch is set for June 2022 and it will reach the Jupiter system on Oct. 2029. 

 An orbital phase around Ganymede is planned (5 months of high elliptical orbit plus 4 months of 500~km circular orbit) after several flybys
of Europa and Callisto (2.6 yr).\\
 The radio link technology will be the same as the one used on MPO (BC) so the range measurements will have a precision of some cm. However, the positioning error of Jupiter with respect to the Earth is expected to be of the order of some meters (during the orbital phase at Ganymede \cite{cappuccio2019}). We assume an RMS of 10m during the Ganymede orbital phase and a conservative value of 100m otherwise.
 
\item VERITAS (Venus Emissivity, Radio Science, InSAR, Topography, and Spectroscopy) is a proposed  NASA Discovery-class mission aimed at the exploration of Venus, for which a 2.7-years (4 Venus cycles) orbital phase is planned.
  The proposed radio link technology will enable the same performances as MPO (BC). We decided to analyze this proposed mission as a placeholder for possible future missions to Venus equipped with state-of-the-art tracking systems. We simulated a 2.7 yrs mission (stating from May 2028) a data spacing of 10 h and an RMS of 4 cm.
  \end{enumerate}

\section{Results and discussion}\label{sec:discussion}
Firstly, we conducted a preliminary check analyzing all considered missions imposing the same set of a-priori (reported in \tref{tabapriori}). This first analysis can be considered as a "base case scenario" with respect to the realistic approach that we will adopt later on. This is intended to assess preliminarly the effectiveness of each mission with respect to the current state of the art.
{In \fref{fig:allmissions} we report, for each parameter, the ratio between the current uncertainty and the RMS attainable under the aforementioned hypotheses. MRO and BC stand out as most performing among the current and the "new" missions respectively. The detailed numerical values used for \fref{fig:allmissions} are reported in \tref{tab:allmissions}.}
\begin{center}
\begin{figure}[h!]
\includegraphics[width=1.0\columnwidth]{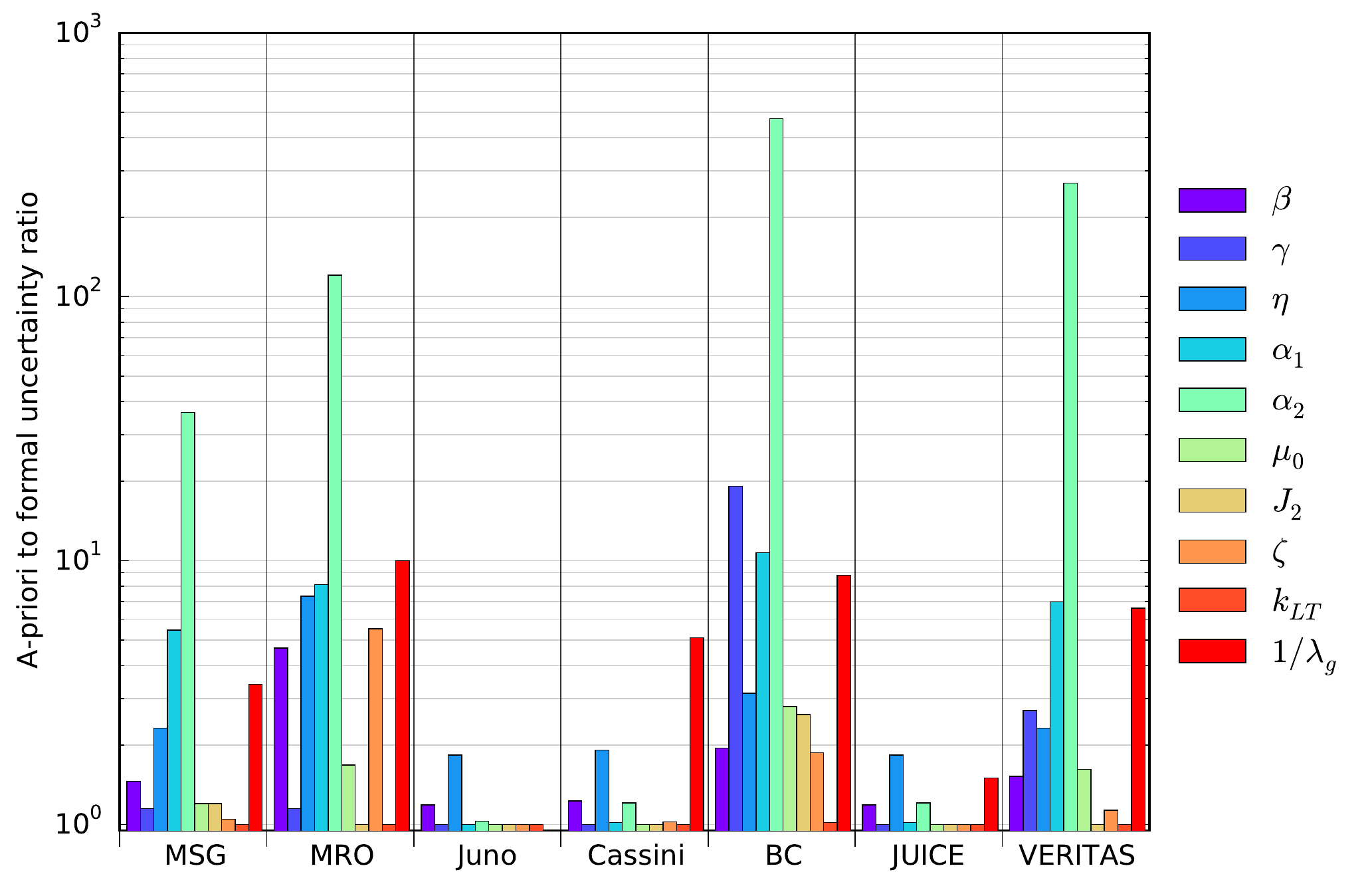}
\caption{\footnotesize Ratio between the current uncertainties of the parameters and their expected RMS, for all missions. Regarding $\lambda_g$ we imposed a reference value of $10^{13}$~km.}
\end{figure}
\label{fig:allmissions}
\end{center}
{ The actual data analysis approach will be different because, in general, new missions always benefit from the results of the previous ones.\\
We simulate this approach in three steps:  first, }
by imposing as a-priori the current knowledge about the parameters 
we get the covariance matrices { for the missions which are currently operative and/or  concluded  MSG, Juno, Cassini,  MRO)}.\\
Afterwards,  taking advantage of the results of these missions we define the new set of a-priori (\tref{tabapriori}, step 2) and we use it for the covariance analysis of  BC.\\
Finally, since  VERITAS (if  confirmed) and JUICE will start to collect data not before year 2028, 
 when the mission BC will be finished, we use the expected results of BC (\tref{tabapriori}, step 3) to define a new set of a-priori (\tref{tabapriori}, step 3) to be considered in the covariance analysis of JUICE/VERITAS.\\
 In all cases we adopt a conservative approach to define the set of a-priori at each step (results have been rounded up). 
 Moreover, we adopt in all cases the same set of a-priori for planets/asteroids ephemerides and GMs even if, in the next 15 years, they will be definitely improved. 
 Results for  MSG, Juno, Cassini and MRO are reported in \tref{tabOld}, while for BC, JUICE and VERITAS are in \tref{tabBJV}.\\
 Comparing the current knowledge (step 1) to the results in \tref{tabOld} one can notice that the data from MRO are the most promising among the current missions:  the improvement, with respect to the current knowledge, spans from a factor 3 ($\mu_0$) to 100 ($\alpha_2$) for all parameters except for $\gamma$ and $k_{LT}$.\\
 Even in the optimistic case (RMS 100m), the results obtained from Cassini data are in general worse than those of MRO.\\
 For $\lambda_g$, our result agrees with the forecast of \cite{will2018} regarding the data analysis of MRO (the lower limit we found is $1.03 \times 10^{14}$~km).\\
 Regarding  MSG, the values we found are in general larger than those reported by \cite{genova2018} since we take into account the uncertainty of the Earth's orbit, and the parameters $\gamma, \alpha_1,\alpha_2$.
For all missions, the contributions to the measurements of the Sun's angular momentum and $J_2$ are negligible.\\
Basing on these results, we build the set of a-priori reported in \tref{tabapriori} (step 2).
\begin{table}[!h]
\caption{Results for the covariance analysis applied to {missions in operation and/or concluded}. A-priori used are reported in \tref{tabapriori} (step 1). We indicate with $\Delta$ the spacing between consecutive simulated normal points expressed  in hours (or days in the case of Juno).} 
\resizebox{\columnwidth}{!}{
\begin{tabular}{l l l l l}
\hline \hline
                                  &   MSG                   & MRO                               &  Juno                                  & Cassini                           \\ 
duration [yr]                     &    4.1                  &    12.0                           &   4.9                                  &  13.2                             \\ 
RMS [m]                           &     1.0                 &      1.0                          & 50.0                                   &       100.0                       \\ 
$\Delta$ [h]                      &      10.0               &    10                             & 53d                                    &    24                             \\ 
$b_{min}$ [$R_\odot$]             &    {     73.7}          &   {     73.7}                     & {     73.7}                            &   {     73.7}                     \\ 
\hline
$\beta$                           &   $4.8 \times 10^{-5}$  &  {       $1.5 \times 10^{-5}$ }   &   {          $5.9 \times 10^{-5}$  }   &  {      $5.7 \times 10^{-5}$   }    \\ 
$\gamma$                          &   $2.0 \times 10^{-5}$  &  {      $2.0 \times 10^{-5}$ }    &   {        $2.3 \times 10^{-5}$  }     &  {     $2.3 \times 10^{-5}$  }      \\ 
$\eta$                            &   $1.9\times 10^{-4}$   &  {        $6.0\times 10^{-5}$  }  &   {       $2.4\times 10^{-4}$   }      &  {      $2.3\times 10^{-4}$   }     \\ 
$\alpha_1$                        &   $1.1 \times 10^{-6}$  &  {       $7.4 \times 10^{-7}$ }   &   {       $6.0 \times 10^{-6}$    }    &  {         $5.9 \times 10^{-6}$   } \\ 
$\alpha_2$                        &   $9.6 \times 10^{-7}$  &  {      $2.9 \times 10^{-7}$ }    &   {        $3.4 \times 10^{-5}$   }    &  {       $2.9 \times 10^{-5}$  }    \\ 
$\mu_0$    [km$^3$ s$^{-2}$]      &   0.35                  &  {        0.25 }                  &   {       0.42 }                       &  {       0.42  }                    \\ 
$J_{2\odot}$                      &   $1.0\times 10^{-8}$   &  {      $1.2\times 10^{-8}$   }   &   {     $1.2\times 10^{-8}$  }         &  {      $1.2\times 10^{-8}$ }       \\ 
$\zeta$  [yr$^{-1}$]              &   $4.1\times 10^{-14}$  &  {       $7.8\times 10^{-15}$ }   &   {      $4.3\times 10^{-14}$}         &  {       $4.2\times 10^{-14}$  }    \\ 
$k_{LT}$                          &   $5.4\times 10^{-3}$   &  {       $5.4\times 10^{-3}$ }    &   {      $5.4\times 10^{-3}$ }         &  {      $5.4\times 10^{-3}$  }      \\ 
$\lambda_g$ [km]                  &   $3.4\times 10^{13}$   &  {      $1.0\times 10^{14}$  }    &   {      $8.3\times 10^{12}$ }         &  {       $5.1\times 10^{13}$ }      \\ 
\hline \hline

 \end{tabular}
 \label{tabOld}
}
\end{table}

\begin{table}[!h]
\begin{ruledtabular}
\caption{Results for the covariance analysis applied to BC (using a-priori reported in \tref{tabapriori}, step 2) and to JUICE/VERITAS (using a-priori, based on BC results, reported in \tref{tabapriori}, step 3). 
We indicate with $\Delta$ the spacing between consecutive simulated normal points. In some cases, since we assumed conservative a-priori (see \tref{tabapriori} step 2), some accuracies 
  are larger than those expected after BC indicated by "$\ast$". This means that in these cases the improvement  is negligible.}
\begin{tabular}{l l l l}
                                  &   BC                         & JUICE                                &  VERITAS                \\        
duration [yr]                     &  2.0                         &    {    2.6 \& 0.8}                      &  2.7                    \\   
RMS [m]                           &  0.04                        &  {100.0 \&  10.0}                         &   0.04                  \\    
$\Delta$ [h]                      &     10                       &    10                                &      10                 \\   
$b_{min}$ [$R_\odot$]             &    7.0                       &   7.0                                &   {    40.0 }           \\  
\hline
$\beta$                           & $1.7 \times 10^{-5}$         &   $1.6 \times 10^{-5}$               &   $1.4 \times 10^{-5}$            \\  
$\gamma$                          & $1.0 \times 10^{-6}$         &   $2.0 \times 10^{-6}$  $\ast$       &   $1.9 \times 10^{-6}$ $\ast$     \\ 
$\eta$                            & $6.9\times 10^{-5}$          &   $6.2\times 10^{-5}$                &   $5.6\times 10^{-5}$             \\ 
$\alpha_1$                        & $3.4 \times 10^{-7}$         &   $5.0 \times 10^{-7}$ $\ast$        &   $4.0 \times 10^{-7}$  $\ast$    \\ 
$\alpha_2$                        & $6.7 \times 10^{-8}$         &   $1.0 \times 10^{-7}$ $\ast$        &   $7.7 \times 10^{-8}$  $\ast$    \\ 
$\mu_0$    [km$^3$ s$^{-2}$]      & 0.08                         &   0.10  $\ast$                       &   0.08                            \\
$J_{2\odot}$                      & $2.8\times 10^{-9}$          &   $4.5\times 10^{-9}$    $\ast$      &   $4.3 \times 10^{-9}$ $\ast$     \\ 
$\zeta$  [yr$^{-1}$]              & $9.2\times 10^{-15}$         &   $1.0\times 10^{-14}$ $\ast$        &   $9.5\times 10^{-15}$ $\ast$     \\ 
$k_{LT}$                          & $5.3\times 10^{-3}$          &   $5.4\times 10^{-3}$ $\ast$         &   $5.4\times 10^{-3}$ $\ast$      \\ 
$\lambda_g$ [km]                  & $1.1\times 10^{14}$          &   $1.0\times 10^{14}$                &   $1.1\times 10^{14}$             \\  
 \end{tabular}
 \label{tabBJV}
\end{ruledtabular}
\end{table}

  Regarding BC, a large improvement (a factor 20) with respect to the results of MRO is expected for the parameter $\gamma$. 
 A factor 2-4 of improvement is expected for $\alpha_1$, $\alpha_2$, $\mu_0$ and $J_{2\odot}$. No significant improvements are expected for the other parameters.\\
 Finally, we found that no improvements of the results of BC are expected after the JUICE and VERITAS range data analysis.
 
\begin{table}[!h]
\begin{ruledtabular}
\caption{A-priori  adopted on the parameters for the covariance analysis in different cases. Step 1: for  missions currently operative and/or finished (VEX, MSG, Juno, Cassini,
 MRO). Step 2: for BC. Step 3: for JUICE and VERITAS.}
\begin{tabular}{llll}
parameter                       & step 1                                          &      step 2                          &      step 3              \\
                               &  (current)                                     & (after MRO, ecc.)                      &     (after BC)         \\      
\hline
  $\beta$                        &  $7.0 \times 10^{-5}$ \cite{imperi2018}   & $3.0 \times 10^{-5}$     &     $2.0 \times 10^{-5}$  \\
  $\gamma$                       &  $2.3 \times 10^{-5}$ \cite{bertotti2003} & $2.0 \times 10^{-5}$     &     $2.0 \times 10^{-6}$  \\
  $\eta$                         &  $4.4 \times 10^{-4}$ \cite{imperi2018}   & $1.0\times 10^{-4}$      &     $1.0\times 10^{-4}$   \\
  $\alpha_1$                     &  $6.0\times 10^{-6}$ \cite{iorio2012}     & $1.0 \times 10^{-6}$     &     $5.0 \times 10^{-7}$  \\
  $\alpha_2$                     &  $3.5\times 10^{-5}$ \cite{iorio2012}     & $5.0 \times 10^{-7}$     &     $1.0 \times 10^{-7}$   \\
  $\mu_0$    [km$^3$ s$^{-2}$]   &   0.42 \footnote{\footnotesize  For the GM of the Sun we adopted 3x the uncertainty reported by \cite{fienga2015}.}              &      0.30                       &      0.10                   \\
  $J_{2\odot}$                   &  $1.2\times 10^{-8}$ \cite{schettino2018}  & $1.1\times 10^{-8}$     &     $4.5\times 10^{-9}$      \\
  $\zeta$  [yr$^{-1}$]           &  $4.3\times 10^{-14}$ \cite{schettino2016} & $1.0\times 10^{-14}$    &     $1.0\times 10^{-14}$    \\
  $k_{LT}$                       &    $5.4\times 10^{-3}$  \cite{park2017}    & $5.4\times 10^{-3}$     &     $5.4\times 10^{-3}$    \\
  $\lambda_g$ [km]               &   none                                     & $9.0\times 10^{13}$     &     $1.0\times 10^{14}$    \\
   \end{tabular}
 \label{tabapriori}
\end{ruledtabular}
\end{table}

\section{Conclusion}\label{sec:conclusion}
In this work we analyzed the possible outcomes in terms of tests of General Relativity, of the radio tracking data analysis of some past, present and future planetary missions. The results we presented have to be intended as a roadmap to guide future data analysis campaigns aiming to set a new and tighter level of accuracy of validation of the General Relativity theory. The approach we employed is based on an updated version of the covariance analysis method, described in \cite{demarchi2016}, which was initially concieved for the BepiColombo Relativity experiment.\\
The approach is fully analytical: all the perturbations on planetary orbits are calculated by solving the Hill's equations.\\
The updating concerns the introduction of the preferred frame parameters ($\alpha_1,\alpha_2$), Eddington parameter $\gamma$, Compton wavelength of the graviton ($\lambda_g$) and Lense-Thirring effect of the Sun.
We also included the  aging of the transponder, the effect of the eccentricity of the Mercury's orbit and the uncertainties of the ephemerides for planets and minor bodies. Finally,
 the code has been extended to deal with the perturbations between two arbitrary bodies orbiting around the Sun.\\
 With these characteristics the code is suitable to compare the sensitivity to each parameter for space missions orbiting around inner or outer planets.\\
The method has been firstly validated by comparing the signatures of the parameters we are interested in with the numerical results for a simulated radio tracking between  Earth and Mercury. {We verified that the analytical model presents minor discrepancies with respect to the validation model, mainly due to the minor neglected effects, that do not yield to significant differences on the results, in terms of formal uncertainties.}
Afterwards, we sequentially performed a covariance analysis to 1) five still operative and/or finished interplanetary missions 2) to BepiColombo and 3) to two missions still to be launched (JUICE) and/or approved (VERITAS). 
Each step {benefits} from the results of the previous one.  We conclude that a significant improvement of the current knowledge can come after the analysis of the range data of MRO. The next important improvement, mainly for the parameter $\gamma$, will be carried out by BC data. We want to stress out  that our simulations are based on a data analysis strategy that considers only one mission at a time, eventually employing as a-priori the results of other missions. It has been preliminarly shown in \cite{cascioli2019} that a strategy based on the simultaneous data analysis of several missions might lead to a further reduction of the formal uncertainties on the parameters of interest via an effective reduction of their correlations. Thus our results represent the first, but fundamental, step towards future combined analysis works. 


\begin{acknowledgments}
The research presented in this work has been carried out at Sapienza University of Rome under a partial sponsorship of the Italian Space Agency within the scope of the contract ASI/2007/I/082/06/0.
The authors would like to aknowledge { A.~Genova for suggesting the relevance of including the mass of the graviton in the analysis, P.~Racioppa for comments that greatly improved \sref{app:constraint},  P.~Cappuccio, V.~Notaro, D.~Durante, L.~Iess for the fruitful discussions about the uncertainty associated with the measurements of JUICE, Juno and Cassini, G.~Schettino and G.~Tommei for the continuous confrontation on the rescaling constraint. } \\

\end{acknowledgments}


\clearpage

\appendix

\onecolumngrid

\section{First order corrections for eccentricity}\label{app:ecc}
 In this Section we report the first-order expansions used in the calculus of the perturbations described in \sref{sec:model1}.\\
All terms into \erefs{eq:da}, (\ref{eq:da2}), (\ref{eq:LT2}), (\ref{eq:gr2}) and (\ref{eq:eta}) can be written using \eref{eq:r0j} and the following formulas (where $n\neq 0$)
\begin{subequations}
\label{eq:appecc}
\begin{eqnarray}
\frac{1}{(r_{0i})^n} & \approx & \frac{1}{(R_{0i})^n} (1+ e_i n \cos \Phi_i);  \label{appa}\\
\frac{\ve r_{0i}}{(r_{0i})^n} &\approx &\frac{1+e_i \cos \Phi_i (n-1) }{R_{0i}^{n-1}}\ve u_r^i+ e_i \frac{2 \sin \Phi_i}{R_{0i}^{n-1}} \ve u_t^i; \label{appb}\\
\dot {\ve r}_{0i} & \approx & n_i R_{0i}  \ve u_r^i+ e_i n_i R_{0i} \left( \sin \Phi_i \ve u_r^i+\cos \Phi_i \ve u_t^i\right); \label{appc}\\
\frac{\ve r_{0i} \cdot  \dot {\ve r}_{0i}}{r_{0i}^3} \dot {\ve r}_{0i} & \approx  & e_i n_i^2  \sin \Phi_i \ve u_t^i; \label{appd}\\
 (\dot r_{0i})^2  \frac{\ve r_{0i} }{r_{0i}^3} & \approx & n_i^2(1+4 e_i   \cos \Phi_i) \ve u_r^i+2 e_i n_i^2 \sin \Phi_i \ve u_t^i; \label{appe}\\
 \frac{\dot {\ve r}_{0i}}{r_{0i}^3} \times \{0,0,1\} & \approx &\frac{n_i}{R_{0i}^2} \left[1+4 e_i \cos \Phi_i \ve u_r^i+e_i \sin \Phi_i \ve u_t^i \right];\label{appf}
\end{eqnarray}
\end{subequations}
Regarding the planet-planet interactions terms, assuming $e_j=0$ for all $j\neq i$, from \eref{eq:r0j} the $i$-to-$j$ vector is
\be
\ve r_{ij}=\ve R_{ij}-e_i  \ve R_i^e
 \ee
where $\ve R_{ij}= R_{0j}\ve u_r^j-R_{0i} \ve u_r^i$ is the same vector in circular approximation. We get
\be
\frac{\ve r_{ij}}{(r_{ij})^n}  \approx \frac{1}{(R_{ij})^n}\left[ \ve R_{ij}  +  e_i \left(\ve R_{i}^e+ n \frac{\ve R_i^e \cdot \ve R_{ij}}{R_{ij}^2} \right) \right]
\label{eq:pl}
\ee
that has been used to evaluate the ($ij$)-term into \eref{eq:r0i} and \eref{eq:eta}.\\
Finally, the quantity $1/(R_{ij})^n$ can be obtained in terms of sinusoidal functions by applying the power-reduction formulae to the well known series expansion (valid for $R_{0i}<R_{0k}$)
\be
\frac{1}{R_{ij}} = \frac{1}{R_{0j}} \sum_{l=0}^\infty \left( \frac{R_{0i}}{R_{0k}} \right)^l P_l (\cos \Phi_{ij})
\label{eq:leg}
\ee
where $P_l$ are the Legendre polynomials.

\section{Solutions for Hill's equations}\label{app:hill}
The perturbations described in this work can be expressed, in the Hill's frame, in terms of sinusoidal and polynomials functions of time, the bigger degree being 3.\\
We define the position/velocity $\{\delta r_i,\delta t_i,\delta \dot r_i,\delta \dot t_i\}$ of the body $i$ in the Hill's frame. The transformation from/to heliocentric coordinates ($\ve r_{0i}, \dot {\ve r}_{0i}$) is
 \be
\begin{split}
  \ve r_{0i} & = (R_{0i} +\delta r_i) \ve u_r^i +   \delta t_i  \ve u_t^i; \\
\dot {\ve  r}_{0i}& =   (\delta \dot r_i  - n_i \delta t_i) \ve u_r^i +  \left[n_i ( R_{0i} + \delta r_i) +\delta \dot t_i\right] \ve u_t^i. \\
  \end{split}
  \label{eq:app1}
\ee

 The equations of motion are
\be
\begin{split}
\delta \ddot  r_i- 2 n_i \delta \dot t_i -3 n_i^2 \delta r_i &= \sum_{i=0}^3 A_{r,i} t^i + \sum_j \left[ C_{r,j} \cos (n_j t) + S_{r,j} \sin(n_j t) \right] + \hat C_r \cos (n_i t)+\hat S_r \sin (n_i t); \\
\delta \ddot  t_i+ 2 n_i \delta \dot r_i                        &=  \sum_{i=0}^3 A_{t,i} t^i + \sum_j \left[ C_{r,j} \cos (n_j t) + S_{t,j} \sin(n_j t) \right] + \hat C_t \cos (n_i t)+\hat S_t \sin (n_i t); \\
\end{split}
\label{eq:hill1}
\ee
where all ($A,S,C,\hat S, \hat C$) coefficients are constants depending on the given perturbation. We indicated with the symbol ($\, \hat{} \, $) the  {\em resonant} terms. Non-resonant frequencies $n_j  \neq n_i $ are arbitrary.\\
 The most general solution is the sum of polynomial ($pol$), resonant ($res$),   non-homogeneous ($nh$) and homogeneous ($h$) terms (we drop the index $i$ for simplicity)

\be
\begin{split}
\delta r &=\delta r_{pol}+\delta r_{res}+ \delta r_{nh}+\delta r_h ; \\
 \delta t &=\delta t_{pol}+\delta t_{res}+ \delta t_{nh}+\delta t_h; \\
  \end{split}
  \label{eq:sol1}
\ee

Solutions for polynomial trends are
\be
\begin{split}
\delta r_{pol} & = {   \cal P}_{R} \cdot \{A_{r0},A_{r1},A_{r2},A_{r3},A_{t0},A_{t1},A_{t2},A_{t3}\}  ;\\
\delta t_{pol} & = {\cal P}_{T} \cdot \{A_{r0},A_{r1},A_{r2},A_{r3},A_{t0},A_{t1},A_{t2},A_{t3}\} ;\\
\end{split}
\ee
where
\onecolumngrid
\be
\begin{split}
{\cal P}_{R} & = \left\{\frac{1}{n^2},\frac{t}{n^2},\frac{n^2 t^2-2}{n^4},\frac{t \left(n^2 t^2-6\right)}{n^4},\frac{2 t}{n},\frac{n^2 t^2-2}{n^3},\frac{2 t \left(n^2 t^2-6\right)}{3
   n^3},\frac{n^4 t^4-12 n^2 t^2+24}{2 n^5}\right\}; \\
{\cal P}_T & =\left\{-\frac{2 t}{n},-\frac{t^2}{n},-\frac{2 t \left(n^2 t^2-6\right)}{3 n^3},\frac{6 t^2}{n^3}-\frac{t^4}{2 n},-\frac{3 t^2}{2},\frac{4 t}{n^2}-\frac{t^3}{2},\frac{4
   t^2}{n^2}-\frac{t^4}{4},-\frac{24 t}{n^4}+\frac{4 t^3}{n^2}-\frac{3 t^5}{20}\right\}.\\
\end{split}
\ee


Resonant contributions are in the form

\be
\begin{split}
\delta r_{res} & = {\cal R}_{R} \cdot  \{\hat S_r, \hat C_r, \hat S_t, \hat C_t \} ;\\
\delta t_{res} & = {\cal R}_{T} \cdot  \{\hat S_r, \hat C_r, \hat S_t, \hat C_t \}  ;\\
\end{split}
\ee

where
\onecolumngrid
\be
\begin{split}
{\cal R}_R & = \left\{   -\frac{t \cos (n t)}{2 n},\frac{t \sin (n t)}{2 n},-\frac{t \sin (n t)}{n},-\frac{t \cos (n t)}{n}    \right\} ;\\
{\cal R}_T & = \left\{\frac{n t \sin (n t)+\cos (n t)}{n^2},\frac{n t \cos (n t)-\sin (n t)}{n^2},\frac{\sin (n t)-2 n t \cos (n t)}{n^2},\frac{2 n t \sin (n t)+\cos (n t)}{n^2}\right\}.\\
\end{split}
\ee


Non-homogeneous solutions are ($n_j \neq n$)

\be
\begin{split}
\delta r_{nh} & =\sum_j {\cal N}_{R,j} \cdot \{S_{r,j},C_{r,j},S_{t,j},C_{t,j}\} ;\\
\delta t_{nh} & = \sum_j {\cal N}_{T,j} \cdot  \{S_{r,j},C_{r,j},S_{t,j},C_{t,j}\}  ;\\
\label{eq:nh}
\end{split}
\ee
where

\be
\begin{split}
{\cal N}_{R,j} & = \left\{\frac{\sin \left(n_j t\right)}{n^2-n_j^2},\frac{\cos \left(n_j t\right)}{n^2-n_j^2},\frac{2 n \cos \left(n_j t \right)}{n_j^3-n^2 n_j},\frac{2 n \sin
   \left(n_j t \right)}{n^2 n_j-n_j^3}\right\};\\
{\cal N}_{T,j} & = \left\{\frac{2 n \cos \left(n_j t \right)}{n^2 n_j-n_j^3},\frac{2 n \sin \left(n_j t \right)}{n_j^3-n^2
   n_j},\frac{\left(n_j^2+3 n^2\right) \sin \left(n_j t \right)}{n_j^2 \left(n^2-n_j^2\right)},\frac{\left(n_j^2+3 n^2\right) \cos \left(n_j t \right)}{n_j^2
   \left(n^2-n_j^2\right)}\right\}.\\
\label{eq:nh2}
\end{split}
\ee

Finally, defining the initial conditions $\{r_0,t_0,\dot r_0,\dot t_0\} = \{\delta r_i(0),\delta t_i(0),\delta \dot r_i(0),\delta \dot t_i(0)\}$, the homogeneous solutions are

\be
\begin{split}
\delta r_h & = -[3 r_0+(2 \dot t_0/n)] \cos (n t)+(\dot r_0/n) \sin (n t) + 4 r_0+2 \dot t_0/n;  \\   
\delta t_h & =  t_0-2 \dot r_0/n -(6 n r_0+3 t_0 ) t +  [6 r_0+(4 \dot t_0/n)] \sin (n t) +(2 \dot r_0/n) \cos(n t).\\
\end{split}
\label{eq:hom2}
\ee

\onecolumngrid
\clearpage
\section{Validation of the code. Comparison with numerical results}\label{app:partials}
Here we report a comparison between the numerical and analytical perturbations on a simulated set of Earth-Mercury range data due to some parameters of our interest. 
 A certain number of effects we neglected are at the origin of the discrepancies ($\approx 15-20$\%), they are  planet-planet interactions (except for the parameter $\eta$ \cite{demarchi2016}), orbital inclinations, second order effects etc. 

\begin{center}
\begin{figure}[h!]
\includegraphics[width=.35\columnwidth]{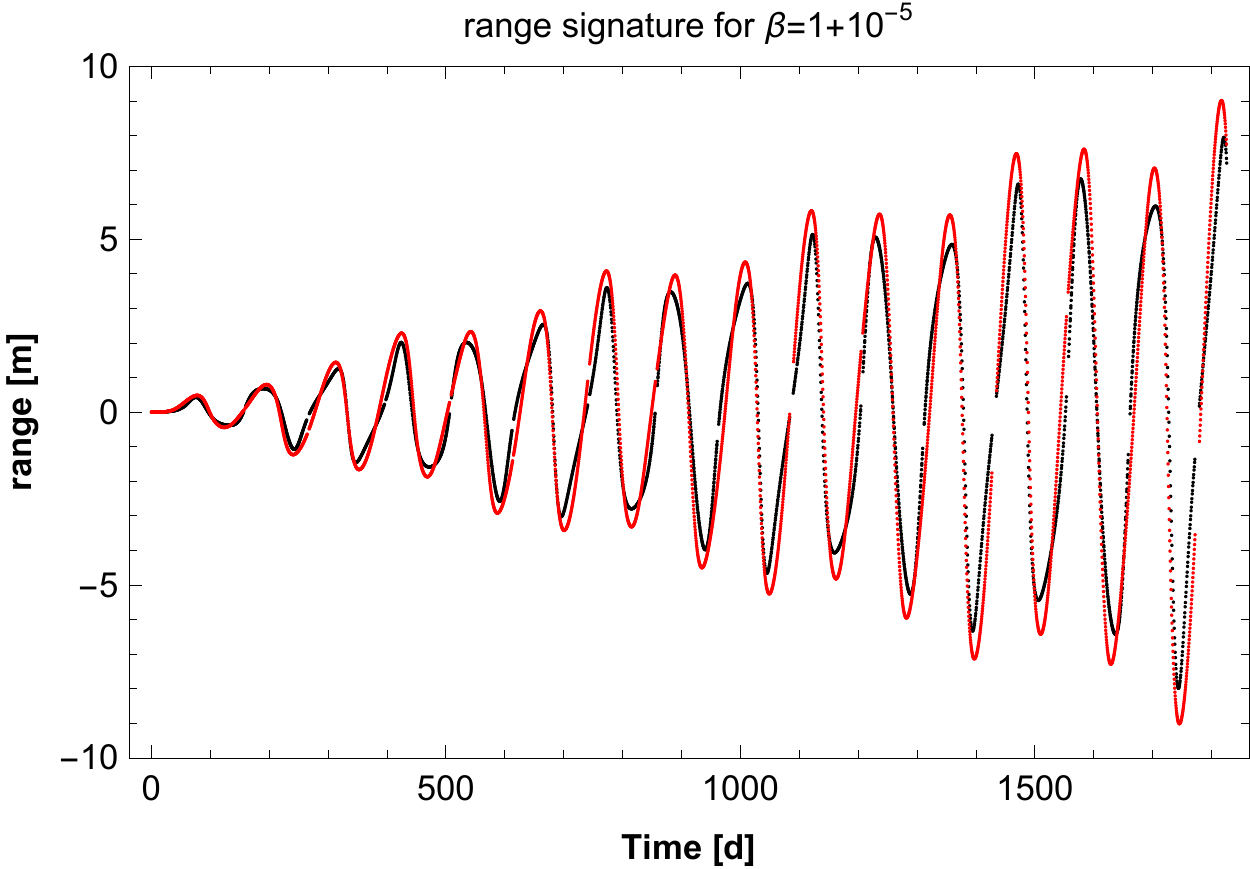}
\includegraphics[width=.35\columnwidth]{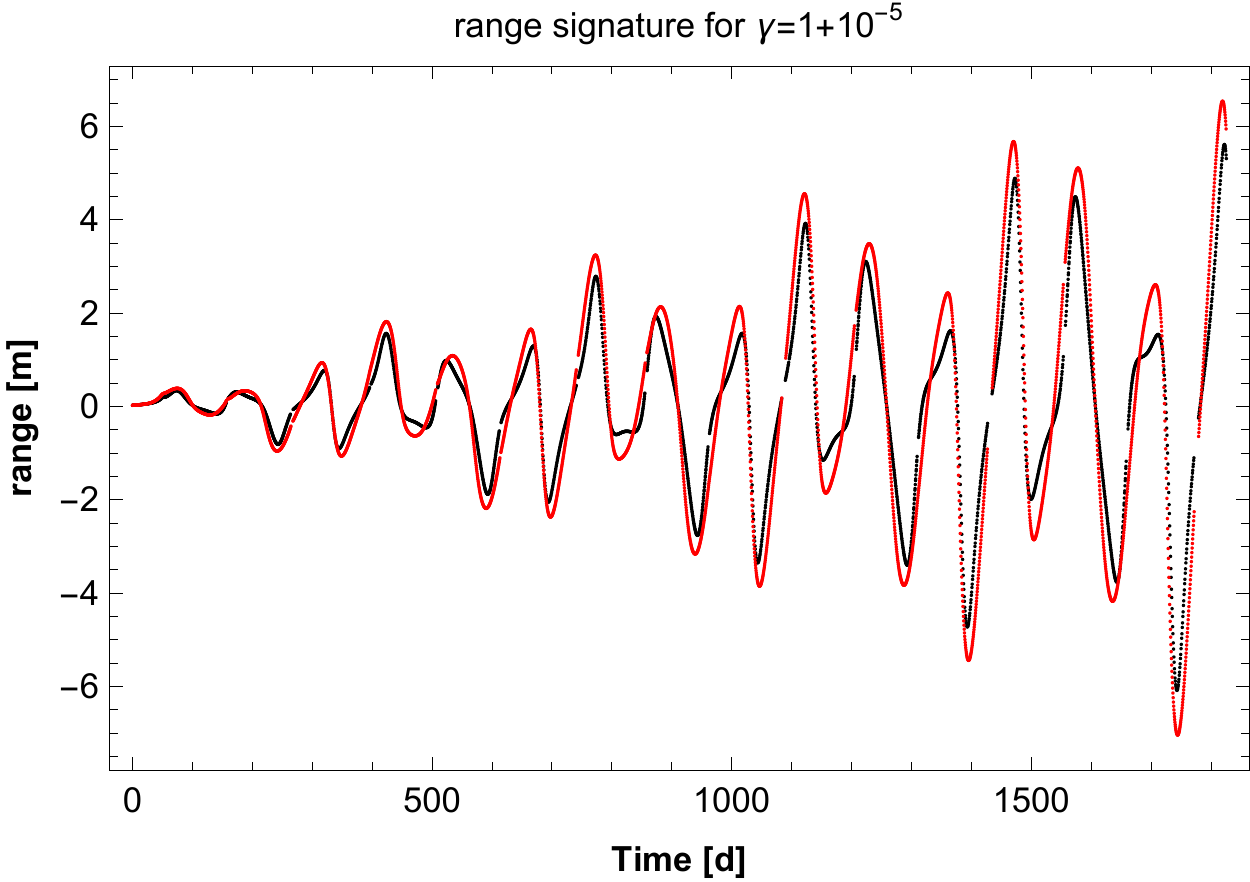}
\includegraphics[width=.35\columnwidth]{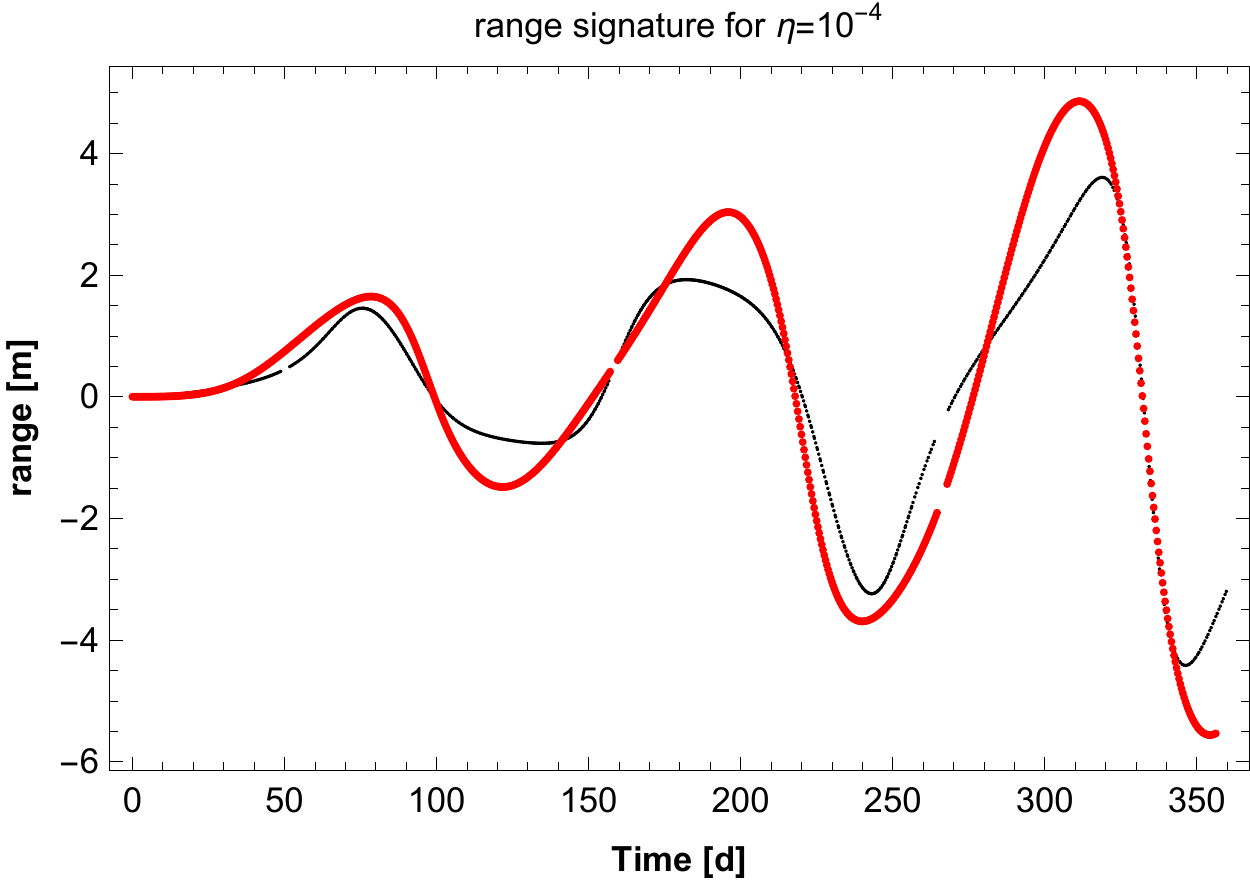}
\includegraphics[width=.35\columnwidth]{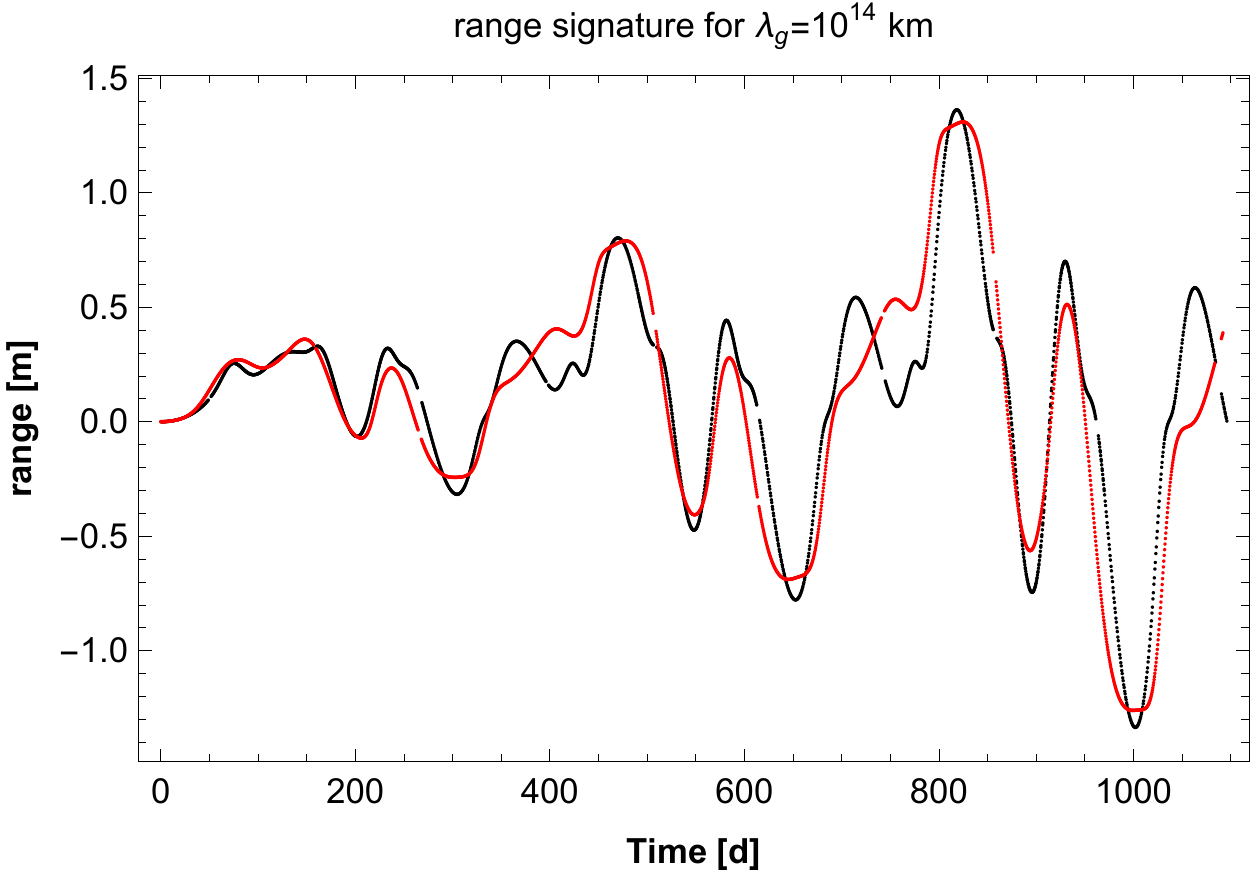}
\includegraphics[width=.35\columnwidth]{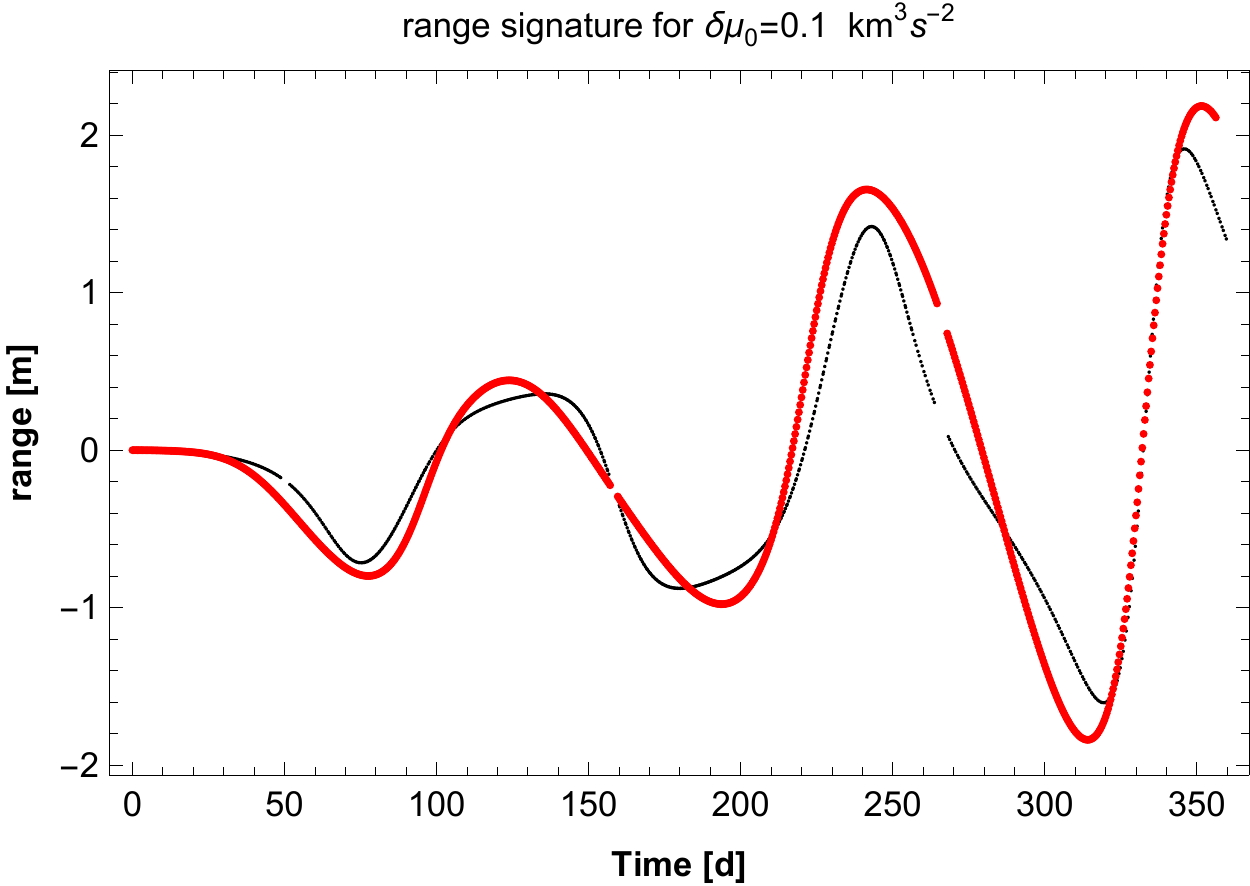}
\includegraphics[width=.35\columnwidth]{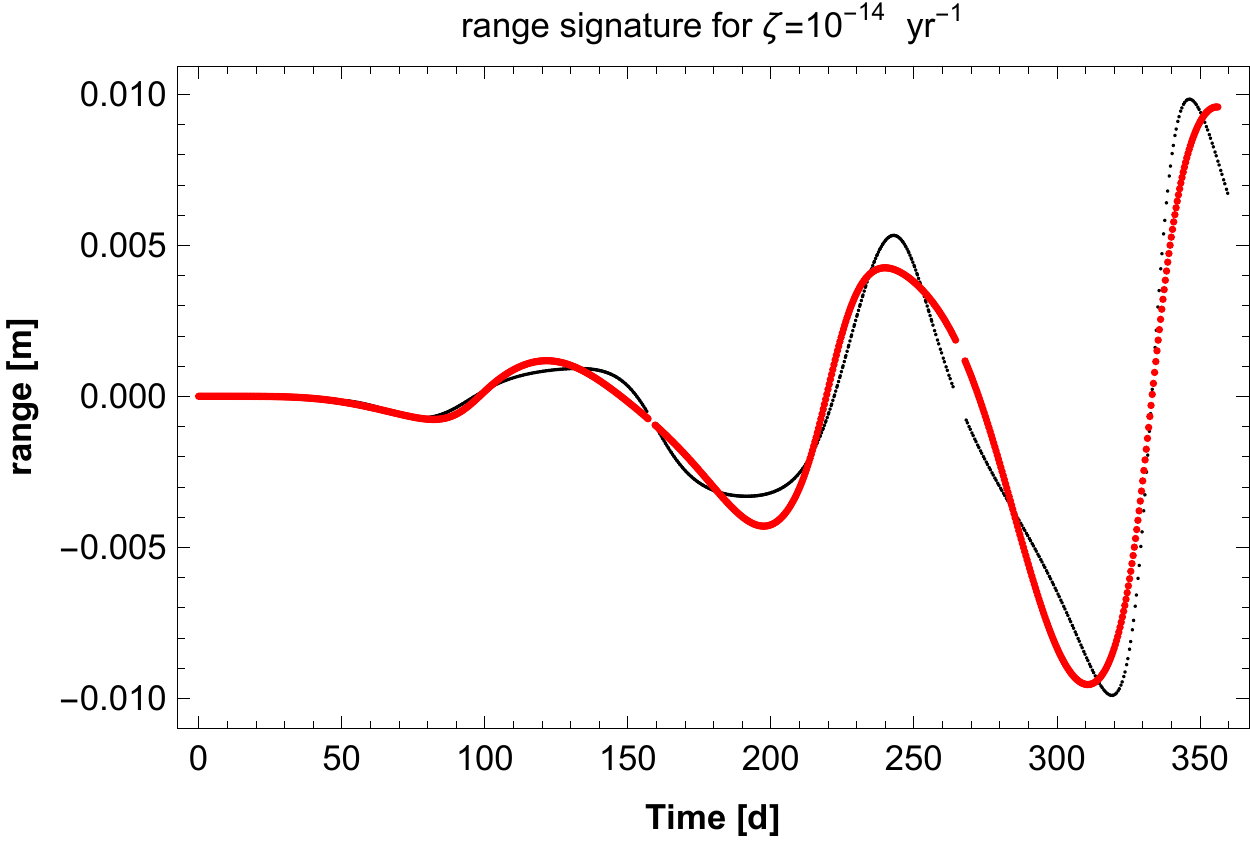}
\includegraphics[width=.35\columnwidth]{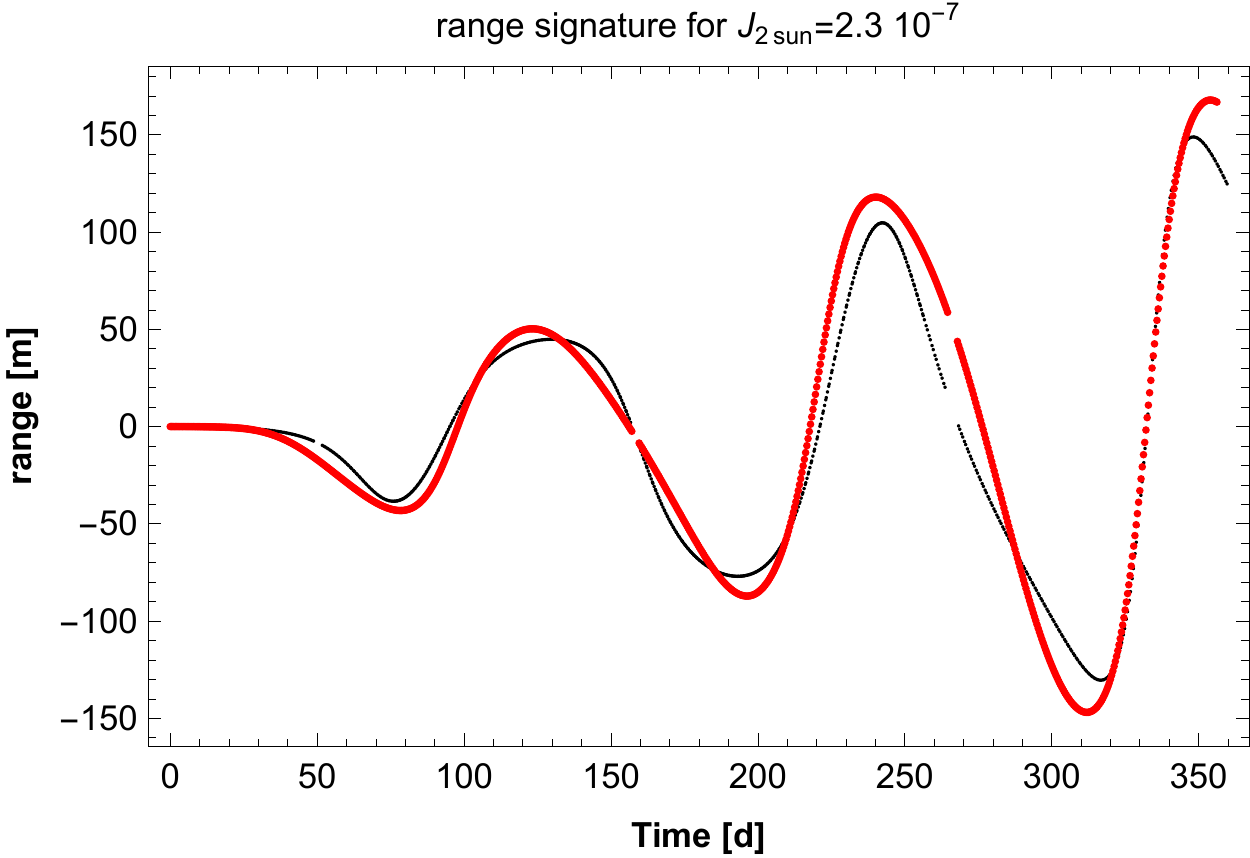}
\includegraphics[width=.35\columnwidth]{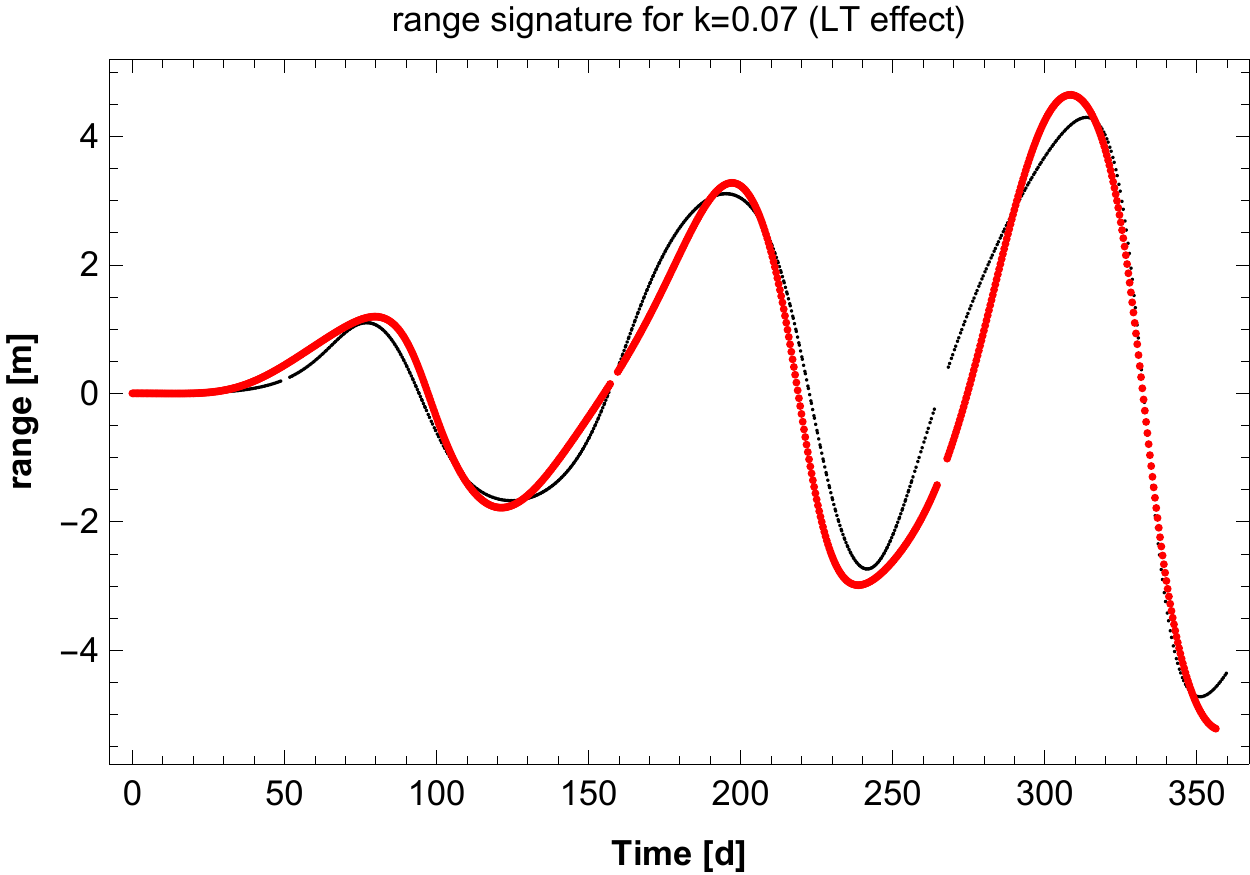}
\caption{\footnotesize Comparison between numerical (black points)  and analytical (red points) Earth-Mercury range signatures for some parameters.}
\end{figure}
\label{fig:partials}
\end{center}

\section{Range signatures for different missions}\label{app:signatures}
 Here we report a collection of the signatures of some parameters for different missions. Regarding $\eta$ we report a comparison between "barycentric" and "heliocentric" signatures (see \sref{sec:ciniz}). In some cases ($\beta$, $\gamma$, $J_{2\odot}$ and $k_{LT}$) the advantage of a mission to Mercury is evident. For $\lambda_g$, on the contrary, the Earth-Jupiter and Earth-Saturn range signals are the largest ones.
\begin{center}
\begin{figure}[h!]
\includegraphics[width=.35\columnwidth]{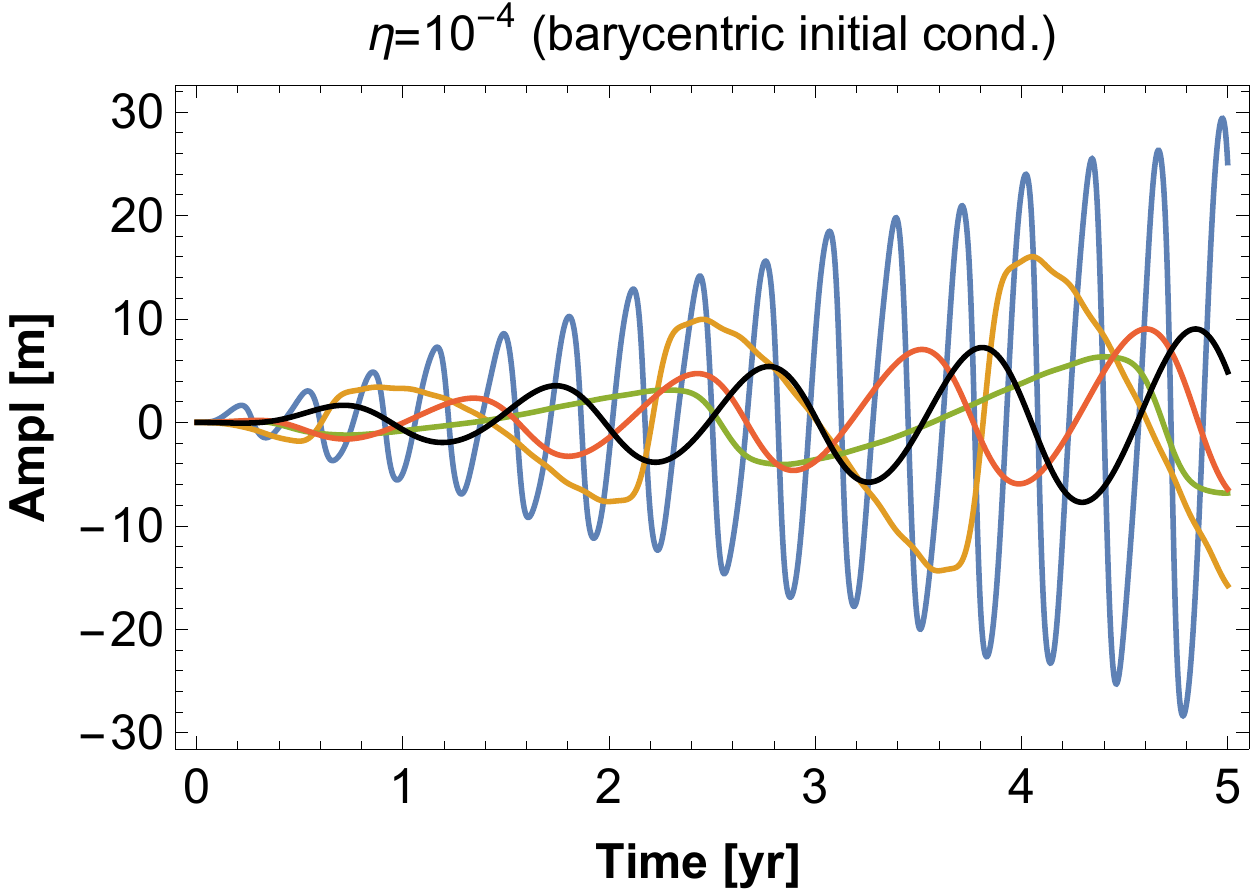}
\includegraphics[width=.35\columnwidth]{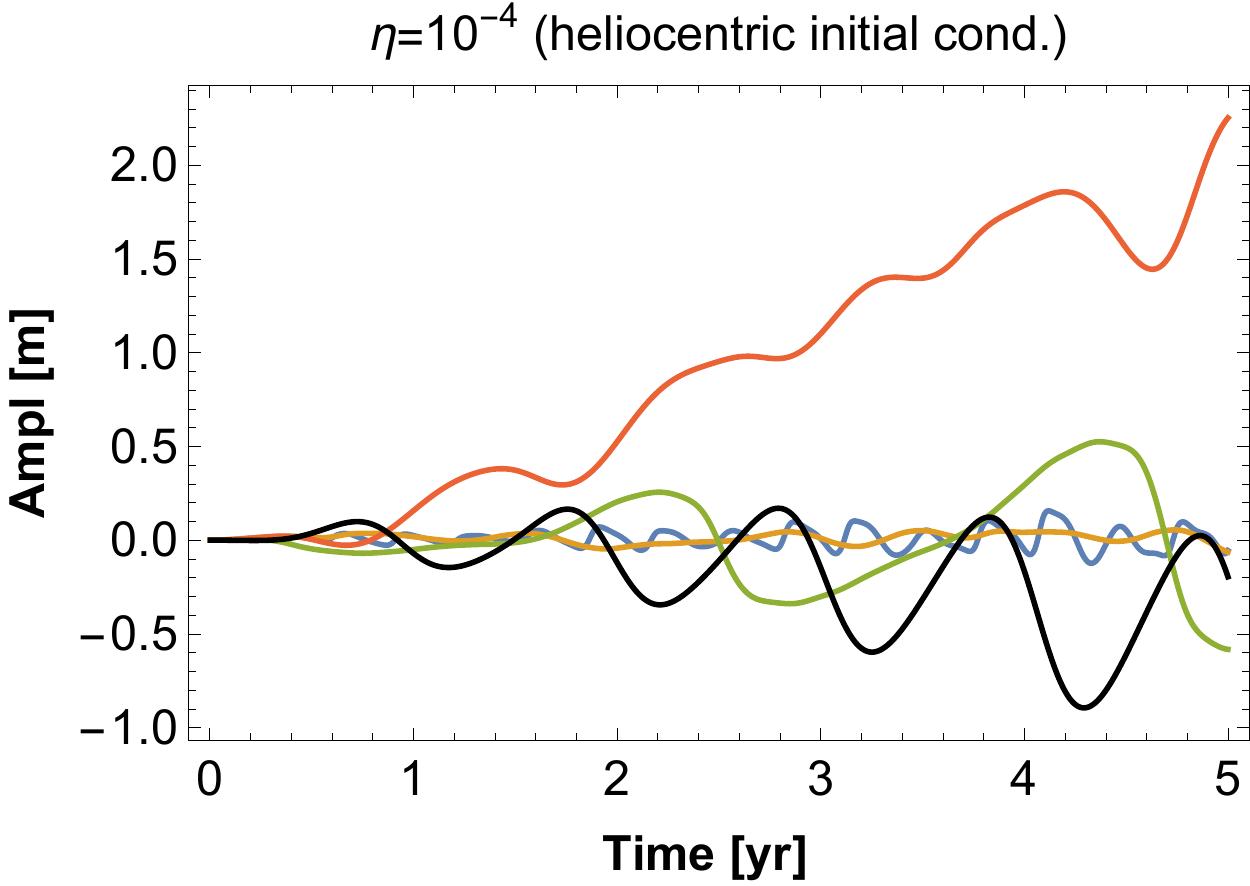}
\includegraphics[width=.35\columnwidth]{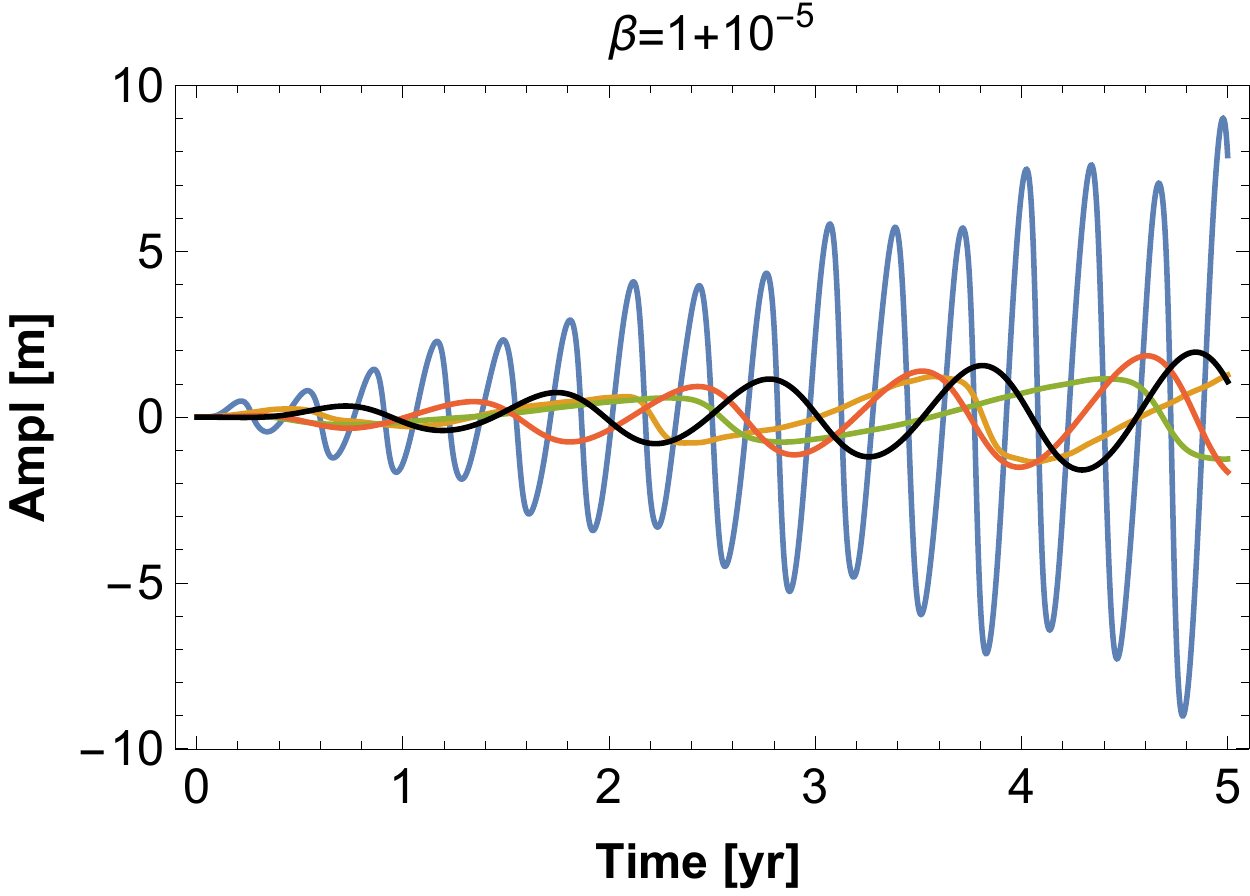}
\includegraphics[width=.35\columnwidth]{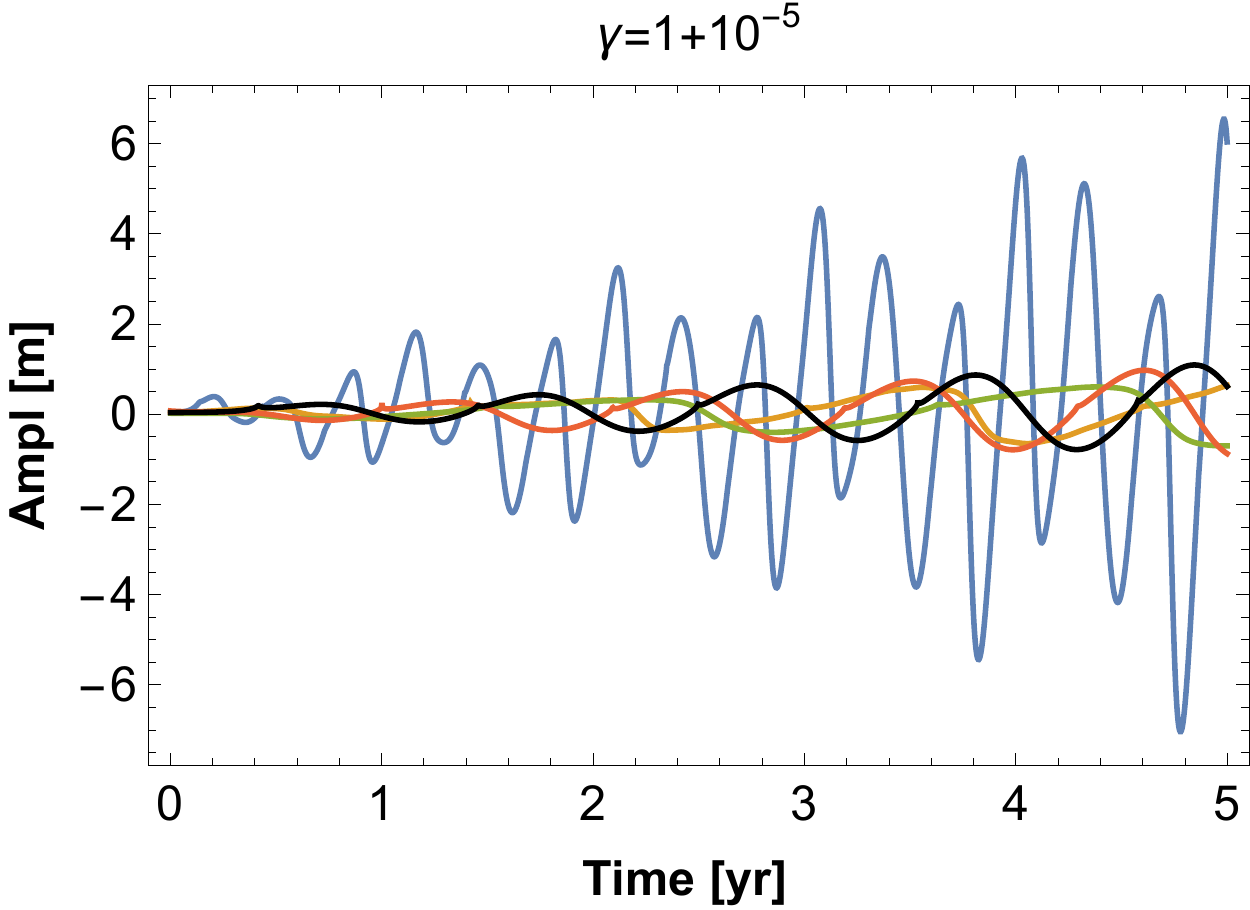}
\includegraphics[width=.35\columnwidth]{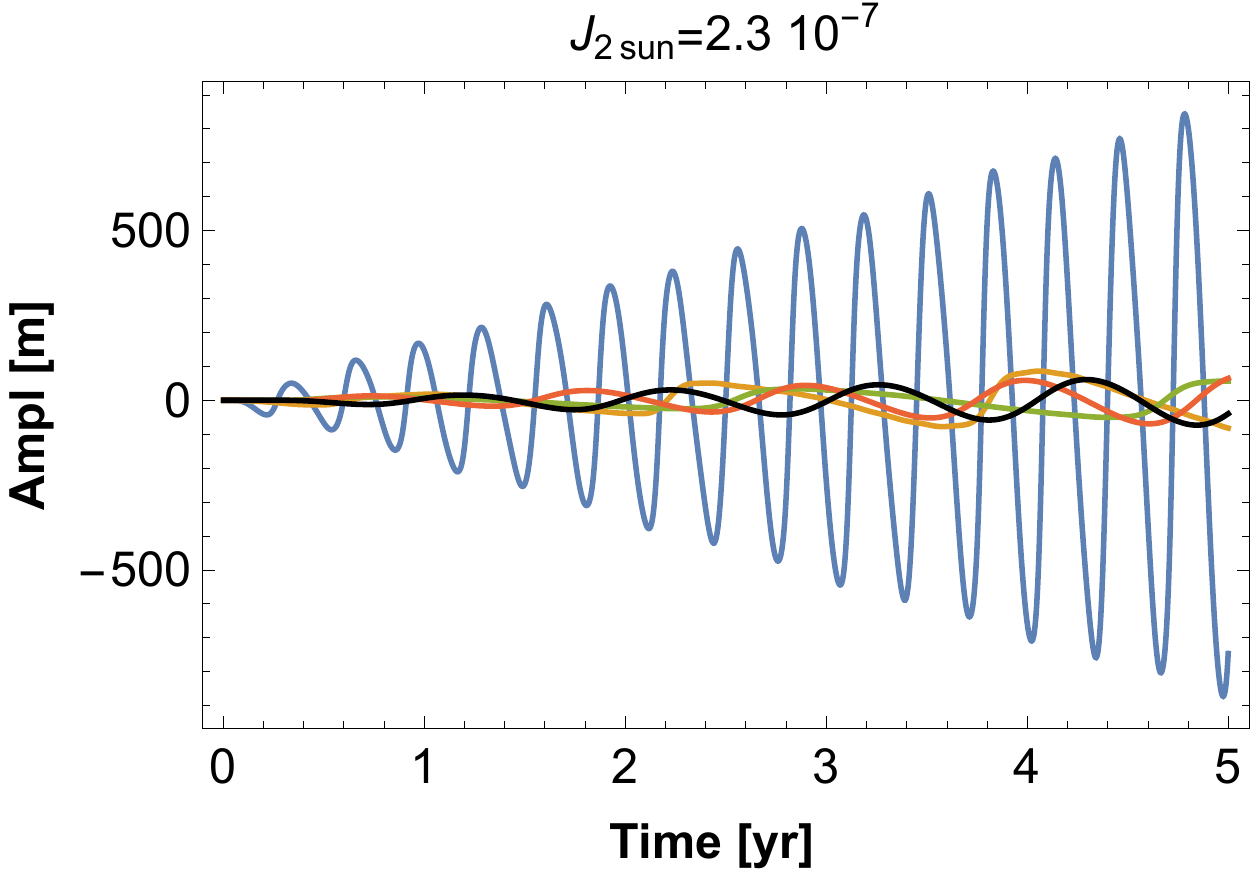}
\includegraphics[width=.35\columnwidth]{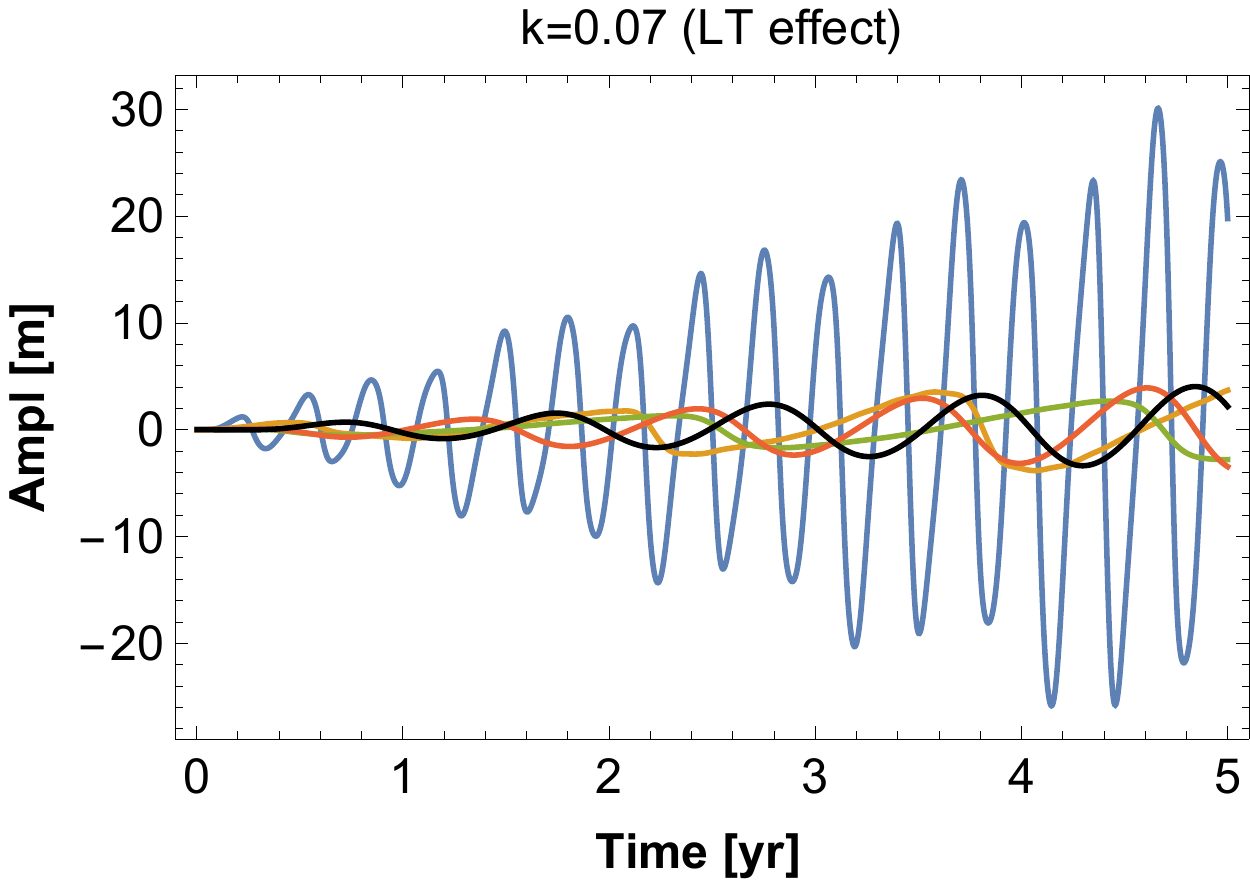}
\includegraphics[width=.35\columnwidth]{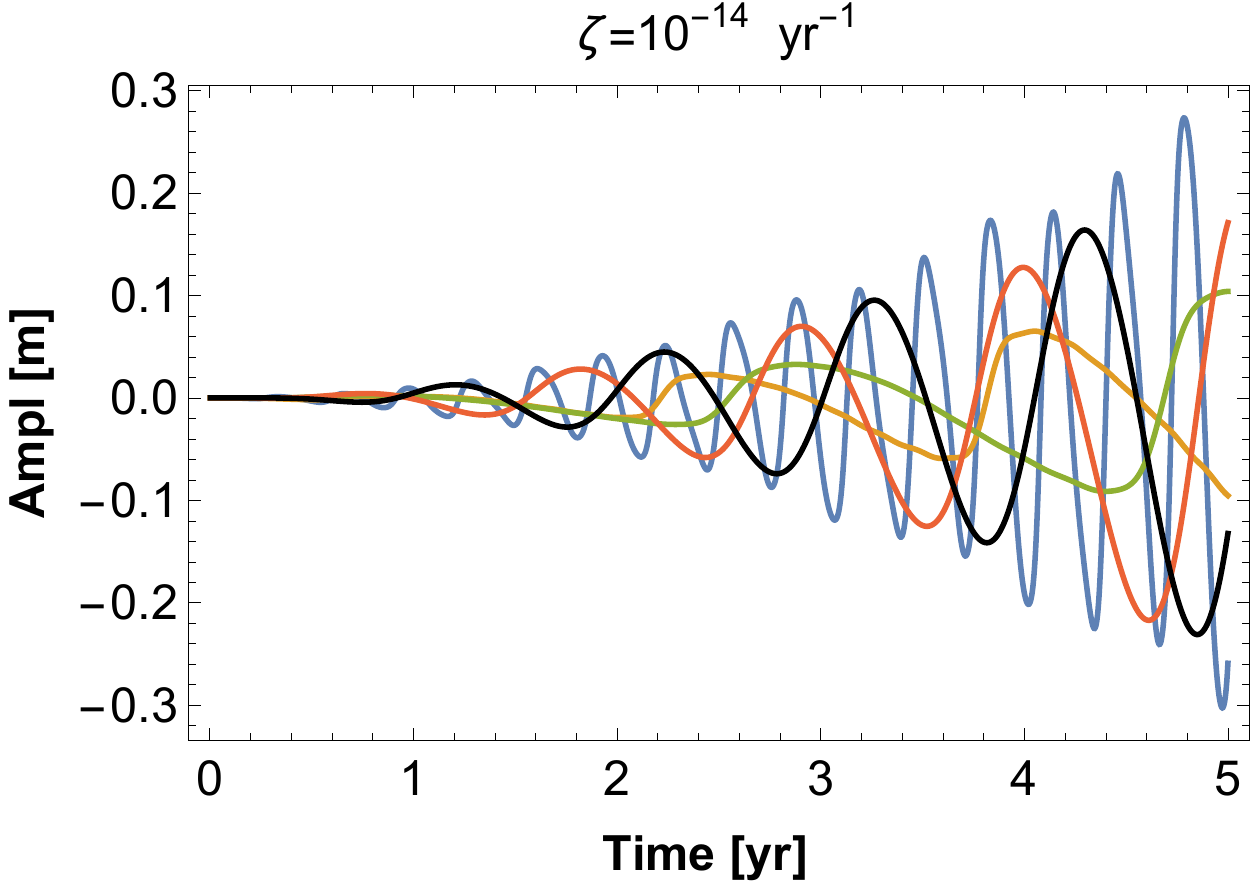}
\includegraphics[width=.35\columnwidth]{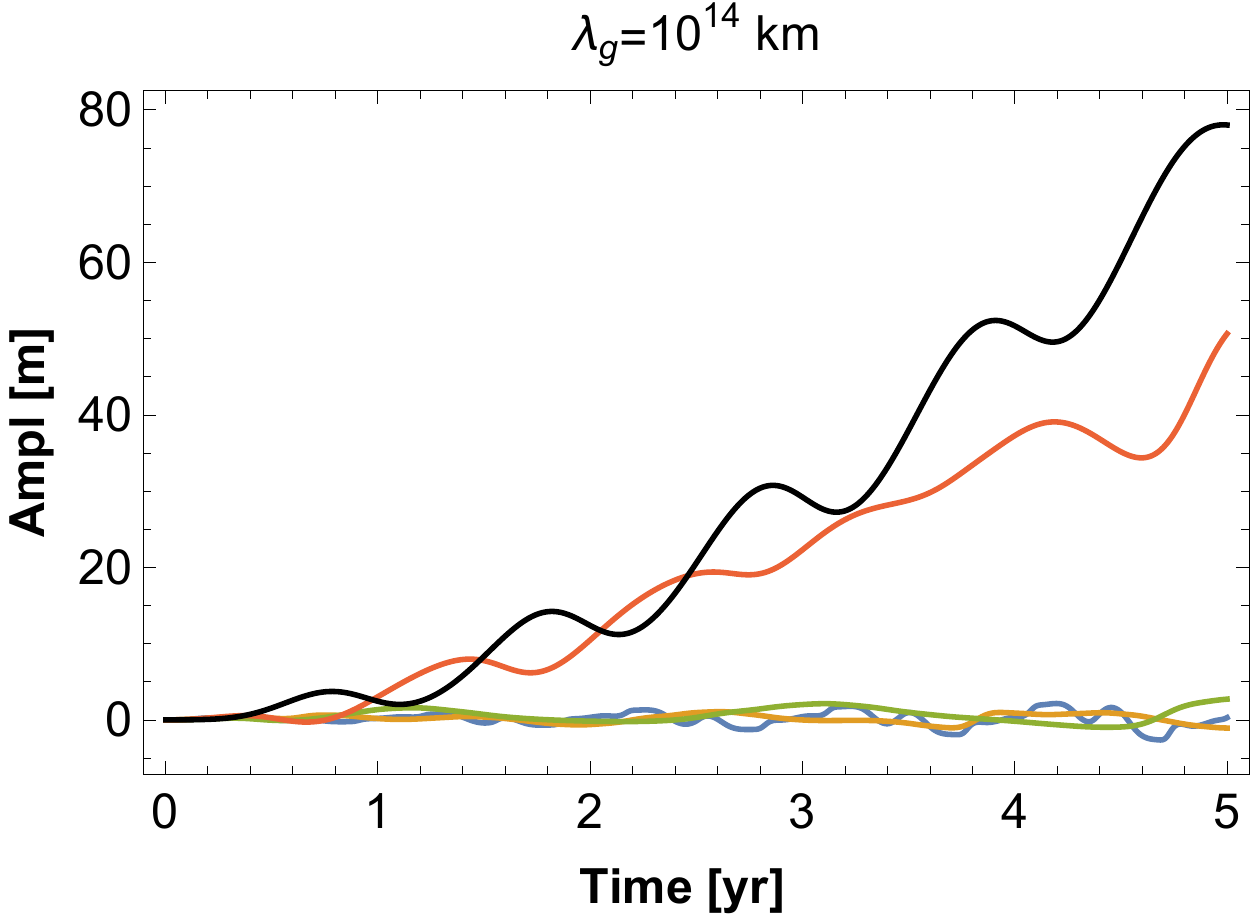}
\caption{\footnotesize Signatures of some parameters for different missions: Earth-Mercury (blue), Earth-Venus (yellow), Earth-Mars (green), Earth-Jupiter (red) and Earth-Saturn (black).}
\end{figure}
\label{fig:signatures}
\end{center}

\section{Performance test for all missions}\label{app:allmissions}
{ For all missions considered in this work, we report the outputs relative to the same set of a-priori (i.e. the current knowledge about the parameters, see \sref{sec:discussion} for details).}

\begin{table}[!h]
\caption{Results for the covariance analysis applied to all missions { with the current uncertainties as a-priori}.} 
\resizebox{\columnwidth}{!}{
\begin{tabular}{l l l l l l l l}
\hline \hline
                                  &   MSG                   & MRO                               &  Juno                                & Cassini                             &  BC                            &  JUICE                       &     VERITAS                   \\ 
duration [yr]                     &    4.1                  &    12.0                           &   4.9                                &  13.2                               &  2.0                           &  2.6 \& 0.8                     &     2.7                       \\ 
RMS [m]                           &     1.0                 &      1.0                          & 50.0                                 &  {    100.0 }                       &  0.04                          &  100.0 \& 10.0                 &     0.04                      \\ 
$\Delta$ [h]                      &      10.0               &    10                             & 53d                                  &    24                               &  10                            &  10                          &     10                        \\ 
$b_{min}$ [$R_\odot$]             &    {     73.7}          &   {     73.7}                     & {73.7}                               &   {     73.7}                       & 7.0                            &  7.0                         &     40.0                      \\ 
\hline
$\beta$                           &   $4.8 \times 10^{-5}$  &  {       $1.5 \times 10^{-5}$ }   &   {        $5.9 \times 10^{-5}$  }   &  {      $5.7 \times 10^{-5}$   }    &  {   $3.6 \times 10^{-5}$ }    &  {   $5.9 \times 10^{-5}$ }  & {   $4.6 \times 10^{-5}$ }    \\ %
$\gamma$                          &   $2.0 \times 10^{-5}$  &  {      $2.0 \times 10^{-5}$ }    &   {      $2.3 \times 10^{-5}$  }     &  {     $2.3 \times 10^{-5}$  }      &  {   $1.2 \times 10^{-6}$ }    &  {   $2.3 \times 10^{-5}$ }  & {   $8.5 \times 10^{-6}$ }    \\ %
$\eta$                            &   $1.9\times 10^{-4}$   &  {        $6.0\times 10^{-5}$  }  &   {     $2.4\times 10^{-4}$   }      &  {      $2.3\times 10^{-4}$   }     &  {   $1.4\times 10^{-4}$  }    &  {   $2.4\times 10^{-4}$  }  & {   $1.9\times 10^{-4}$  }    \\ %
$\alpha_1$                        &   $1.1 \times 10^{-6}$  &  {       $7.4 \times 10^{-7}$ }   &   {     $6.0 \times 10^{-6}$    }    &  {         $5.9 \times 10^{-6}$   } &  {   $5.6 \times 10^{-7}$ }    &  {   $5.9 \times 10^{-6}$ }  & {   $8.6 \times 10^{-7}$ }    \\ %
$\alpha_2$                        &   $9.6 \times 10^{-7}$  &  {      $2.9 \times 10^{-7}$ }    &   {      $3.4 \times 10^{-5}$   }    &  {       $2.9 \times 10^{-5}$  }    &  {   $7.4 \times 10^{-8}$ }    &  {   $2.9 \times 10^{-5}$ }  & {   $1.3 \times 10^{-7}$ }    \\ %
$\mu_0$    [km$^3$ s$^{-2}$]      &   0.35                  &  {        0.25 }                  &   {     0.42 }                       &  {       0.42  }                    &  {   0.15 }                    &  {   0.42 }                  & {   0.26 }                    \\ %
$J_{2\odot}$                      &   $1.0\times 10^{-8}$   &  {      $1.2\times 10^{-8}$   }   &   {   $1.2\times 10^{-8}$  }         &  {      $1.2\times 10^{-8}$ }       &  {   $4.6\times 10^{-9}$   }   &  {   $1.2\times 10^{-8}$   } & {   $1.2\times 10^{-8}$   }   \\ %
$\zeta$  [yr$^{-1}$]              &   $4.1\times 10^{-14}$  &  {       $7.8\times 10^{-15}$ }   &   {    $4.3\times 10^{-14}$}         &  {       $4.2\times 10^{-14}$  }    &  {   $2.3\times 10^{-14}$ }    &  {   $4.3\times 10^{-14}$ }  & {   $3.8\times 10^{-14}$ }    \\ %
$k_{LT}$                          &   $5.4\times 10^{-3}$   &  {       $5.4\times 10^{-3}$ }    &   {    $5.4\times 10^{-3}$ }         &  {      $5.4\times 10^{-3}$  }      &  {   $5.3\times 10^{-3}$ }     &  {   $5.4\times 10^{-3}$ }   & {   $5.4\times 10^{-3}$ }     \\ %
$\lambda_g$ [km]                  &   $3.4\times 10^{13}$   &  {      $1.0\times 10^{14}$  }    &   {    $8.3\times 10^{12}$ }         &  {       $5.1\times 10^{13}$ }      &  {   $8.8\times 10^{13}$  }    &  {   $1.5\times 10^{13}$  }  & {   $6.6\times 10^{13}$  }    \\ %
\hline \hline

 \end{tabular}
 \label{tab:allmissions}
}
\end{table}

\section{Constraint Proof}\label{app:constraint}
In this section we prove that in a minimum variance least squares (MVLS) orbit determination problem, in which linear constraints on the solve-for parameters are applied, the covariance matrix is independent of the epoch at which the constraints are specified.\\
The general solution of the MVLS can be written as
\begin{equation}
\hat{\textbf{x}}_{0} = \left( \ve H^T \ve W \ve H\right)^{-1} \ve H^T \ve W \textbf{y}
\end{equation}
where $\hat{\textbf{x}}_{0}$ is the vector of differential corrections to the solve-for parameters computed at epoch $t=t_0$, $\ve H$ is the design matrix, $\ve W$ is the weight matrix (that in the MVLS corresponds to the inverse of the observation noise covariance matrix) and $\textbf{y}$ is the observation deviation vector (for the derivation of the solution see \cite{tapley2004}). Note that we did not include a-priori informations in the solution for a matter of simplicity, but their inclusion would be straightforward.
One or more linear constraints (as for example fixing the X coordinate of the Earth at a certain epoch) can be introduced in the problem and the easiest way to factor them in is to treat them as additional observations. \\ We can write a generic linear constraint as
$$ c_0= U(t)-M(t)=\mathcal{N}(0,\sigma^2_C)$$
where $U$ is the parameter that is being constrained, $M$ is its constrained value and $\mathcal{N}(0,\sigma^2_C)$ denotes a normally distributed random variable with 0 mean and variance $\sigma^2_C$. \\ 
If the constraint is computed at the estimation epoch (i.e. at $t=t_0$) the design matrix when the constraint is factored in becomes:
\begin{equation}
\ve H = \begin{bmatrix}
 \bar{\ve H}\\
\dfrac{\partial c_0}{\partial \textbf{x}_0} 
\end{bmatrix} = \begin{bmatrix}
\bar{\ve H}\\
\textbf{u}^T

\end{bmatrix}
\end{equation}
where $\bar{\ve H}$ is the partition of $\ve H$ relative to the actual radiometric observations and $\frac{\partial c_0}{\partial \textbf{x}_0}=\textbf{u}^T $ is the additional row due to the constraint.$\textbf{u} $ corresponds to a null column vector except for the index corresponding to the parameter that is being constrained (i.e. $U$). The weights matrix can be in the same way partitioned as:
\begin{equation}
\ve W=
\left[
\begin{array}{c|c}
\bar{\ve W} & 0 \\
\hline
0 & \sigma^{-2}_C
\end{array}
\right]
\end{equation}
Thus the inverse of the covariance matrix $P^{-1}$ can be written as:
\begin{equation}
  \ve P^{-1} = \ve H^T \ve W \ve H = \bar{\ve H}\bar{\ve W}\bar{\ve H} + \frac{1}{\sigma^2_C}\textbf{u}\textbf{u}^T
\end{equation}

In the case the constraint is given at a different time ($t=t_1$):

$$ c_1= U(t_1)-M(t_1) = \mathcal{N}(0,\sigma^2_{C_{1}})$$

Where this time, in order for the constraint to provide the same amount of information as the one treated before, also the weight must be mapped at the correct epoch, thus:
\begin{equation}
\label{eq:sigmaMapping}
 \sigma^2_{C_{1}} = \Phi(t_1,t_0)\sigma^2_C\Phi^T(t_1,t_0) 
 \end{equation}
Where $\Phi(t_1,t_0)$ is the state transition matrix that maps the state from $t_0$ to $t_1$ ( for a complete discussion about the state transition matrix refer to \cite{tapley2004}).\\
In this case the mapping matrix (referred always to the estimation epoch $t_0$):
\begin{equation}
\ve H = \begin{bmatrix}
\bar{H}\\
\dfrac{\partial c_1}{\partial \ve x_0} 

\end{bmatrix}= \begin{bmatrix}
\bar{H}\\
\dfrac{\partial c_0}{\partial \ve x_0}\Phi(t_1,t_0) 

\end{bmatrix} = \begin{bmatrix}
\bar{H}\\
\ve u^T \Phi(t_1,t_0)

\end{bmatrix}
\end{equation}
Then the covariance matrix:
\begin{equation}
\label{eq:invcovariance}
 \ve P^{-1} = \ve H^T \ve W \ve H = \bar{\ve H}\bar{\ve W}\bar{\ve H} + \textbf{u}\Phi^T(t_1,t_0)\frac{1}{\sigma^2_{C_1}}\Phi(t_1,t_0)\textbf{u}^T
 \end{equation}

Using that (from \ref{eq:sigmaMapping}):
\begin{equation}
\label{eq:sigmaMapping2}
\sigma_{C_{1}}^{-2} = \Phi^{-T}(t_1,t_0)\frac{1}{\sigma^2_C}\Phi^{-1}(t_1,t_0)
\end{equation}

and substituting \ref{eq:sigmaMapping2} in \ref{eq:invcovariance}
\begin{equation}
   \ve P^{-1} = \bar{\ve H}\bar{\ve W}\bar{\ve H} + \textbf{u}\Phi^T(t_1,t_0)\Phi^{-T}(t_1,t_0)\frac{1}{\sigma^2_C}\Phi^{-1}(t_1,t_0)\Phi(t_1,t_0)\textbf{u}^T = \bar{\ve H}\bar{\ve W}\bar{\ve H} +\frac{1}{\sigma^2_C}\textbf{u}\textbf{u}^T
\end{equation}
We proved that if the a-priori uncertainties on the constraints are properly mapped the inverse of the covariance matrix is unchanged. We can conclude that also the covariance matrices will remain unchanged.

\bibliography{Rel_Exp_many_probes_refs}

\end{document}